\documentclass[aps,pre,groupedaddress,showpacs,showkeys]{revtex4}
\usepackage{epsfig,graphicx}

\begin{document}

\title[Supersymmetric Extensions of Calogero--Moser--Sutherland like Models ...]
       {Supersymmetric Extensions of Calogero--Moser--Sutherland like Models:
        Construction and Some Solutions}

\author{Heiner Kohler\dag\ and Thomas Guhr\ddag}
\affiliation{\dag\
         Institut f\"{u}r Theoretische Physik,
         Philosophenweg 19, Universit\"{a}t Heidelberg, Germany}
\affiliation{\ddag\
         Matematisk Fysik, LTH, Lunds Universitet,
         Box 118, 22100 Lund, Sweden}

\begin{abstract}

  We introduce a new class of models for interacting particles.  Our
  construction is based on Jacobians for the radial coordinates on
  certain superspaces. The resulting models contain two parameters
  determining the strengths of the interactions.  This extends and
  generalizes the models of the Calogero--Moser--Sutherland type for
  interacting particles in ordinary spaces. The latter ones are
  included in our models as special cases.  Using results which we
  obtained previously for spherical functions in superspaces, we
  obtain various properties and some explicit forms for the solutions.
  We present physical interpretations. Our models involve two kinds of
  interacting particles. One of the models can be viewed as describing
  interacting electrons in a lower and upper band of a
  one--dimensional semiconductor. Another model is
  quasi--two--dimensional. Two kinds of particles are confined to two
  different spatial directions, the interaction contains
  dipole--dipole or tensor forces.

\end{abstract}


\maketitle

\section{Introduction}
\label{sec1}

There is an intimate relation between group theory and certain
one--dimensional exactly solvable
systems~\cite{HC,HC1,gel50,ber57,HEL}. The radial part of the
Laplace--Beltrami operator on symmetric spaces induces in a natural
way an interacting one--dimensional many--body Hamiltonian with a
characteristic $g v^{-2}(x_n-x_m)$ interaction between the particles
at positions $x_n$ and $x_m$. Here, $g$ is the coupling constant and
the function $v$ may be a sine, a hyperbolic sine or the identity,
depending on the curvature of the symmetric space under consideration.
These and similar systems have been studied first by Calogero and
Sutherland~\cite{cal69,cal71,sut72}. They have much in common with the
Brownian motion model studied by Dyson as early as in
1962~\cite{DYS1,DYS2}. Other forms of the potential have been
introduced, such as the Toda lattice~\cite{gut80a,gut80b} or the
Weierstrass function, which generalizes the original form of
interaction. We refer to all models as Calogero--Moser--Sutherland
(CMS) models irrespectively of the interaction potential and the
underlying Lie algebra.

The first proof of exact integrability of some CMS--Hamiltonians have
been given in \cite{OP77}. Later a more general proof has been given
in \cite{heck87,heck91} by very different arguments. In this context,
we also refer to the work in Ref.~\cite{etin95}.

More recently, these models have been studied in the framework of
supersymmetric quantum mechanics~\cite{susy1,susy2}. Although we work
with supersymmetry as well, our approach is different from this.
Generalizations to higher space dimensions~\cite{gos97,kah98,MEL04b}
have also been proposed. Extensive reviews are given in
Refs.~\cite{OP,CAL}.

Our supersymmetric construction extends and generalizes the group
theoretical approach in ordinary spaces by exploring the relation
between the radial part of Laplace operators on symmetric superspaces
and certain Schr\"odinger operators: In some cases, i.e.~for special
values of the coupling constant $g$, the solution of the interacting
particle Hamiltonian can be written as an integral over the classical
matrix groups, the orthogonal, the unitary and the symplectic group.
These groups are labeled by the Dyson index $\beta=1,2,4$,
respectively. The coupling constant $g$ is a function of the Dyson
index $\beta$.  Similar relations for Schr\"odinger operators exist
also in superspace~\cite{TG,GUH4,GUKOP2,GUKO2}. A classification of
matrix supergroups and more general of symmetric superspaces has been
given in Ref.~\cite{MRZ1}. In this contribution, we introduce a
labeling of symmetric superspaces in terms of a pair of numbers
$(\beta_1,\beta_2)$ akin to Dyson's index $\beta$ in ordinary space.
This label may further be continued analytically in $\beta_1$ and
$\beta_2$ to arbitrary combinations $(\beta_1,\beta_2)$.  Our
construction leads to a natural supersymmetric generalization of the
CMS model for interacting particles. Hence, we arrive at a new class
of many-body systems.  They are likely to be exactly solvable in the
allowed parameter region.

Our construction goes considerably beyond the one by Sergeev and
Veselov~\cite{ser,serves1,serves2}. These authors arrived at
superanalogues of CMS models, starting from the underlying root spaces
of the superalgebra.  They also give a solution in terms of
superanalogues of Jack polynomials. Their models however, depend only
on one parameter and are therefore different from ours which crucially
depend on two. Some of our models are related to the many species
generalization of CMS models in Refs.~\cite{MEL03,MEL04a}.  In
contrast to our approach, the latter construction is ad hoc and it is
not based on superspaces.

The models we are investigating have been communicated in
\cite{guk05}, where emphasis was put on their interpretation and
possible applications. Here we focus on mathematical aspects of the
models. In particular the question of exact solvablity is discussed
and exact solutions for certain parameters $\beta_1$, $\beta_2$ are
presented.

The paper is organized as follows: For the convenience of the reader
we briefly compile some results for the models for interacting
particles in ordinary space in Section~\ref{sec2}. Various
supersymmetric generalizations of the models for interacting particles
are presented in Section~\ref{sec3}. In Section~\ref{sec41}, we find
certain solutions by deriving a new recursion formula. In
Section~\ref{sec4}, we give an extensive interpretation of the
physical systems described by the supersymmetric models. A brief
version of this section can be found in \cite{guk05}. We summarize and
conclude in Section~\ref{sec5}.

\section{Models for Interacting Particles in Ordinary Space}
\label{sec2}

In Section~\ref{sec21}, we sketch the connection between ordinary
groups and the many--particle Hamiltonian. We discuss the connection
to the recursion formula in Section~\ref{sec22}.

\subsection{Differential Equation and its Interpretation}
\label{sec21}

The connection between some models of the CMS type in ordinary space
and some radial Laplaceans appearing in group theory~\cite{OP} is seen
by considering the eigenvalue equation
\begin{equation}
  \Delta_x^{(\beta)} \Phi_N^{(\beta)}(x,k)
  = -\left(\sum_{n=1}^Nk_n^2\right) \Phi_N^{(\beta)}(x,k) \ .
  \label{rado}
\end{equation}
The $N$ variables $x_n, \ n=1,\ldots,N$ are interpreted as the
positions of the particles later on. There is a further set of $N$
variables $k_n, \ n=1,\ldots,N$ which will play the r\^ole of quantum
numbers. The operator $\Delta_x$ depends on a parameter $\beta$ and is
given by
\begin{equation}
\Delta_x^{(\beta)} = \sum_{n=1}^N \frac{1}{|\Delta_N(x)|^\beta}
\frac{\partial}{\partial x_n}|\Delta_N(x)|^\beta
\frac{\partial}{\partial x_n} \ ,
\label{opo}
\end{equation}
where
\begin{equation}
  \Delta_N(x) = \prod_{n<m}(x_n-x_m) \label{vd}
\end{equation}
is the Vandermonde determinant. If the symmetry condition
$\Phi_N^{(\beta)}(x,k)=\Phi_N^{(\beta)}(k,x)$ and the initial
condition $\Phi_N^{(\beta)}(0,k)=1$ are required, the solution of the
eigenvalue equation~(\ref{rado}) is for $\beta=1,2,4$ equivalent to
group integrals over ${\rm O}(N)$, ${\rm U}(N)$ and ${\rm USp}(2N)$,
respectively. These integrals are referred to as spherical
functions~\cite{HEL1}. We notice that they are different from the
group integral which Harish--Chandra investigated in
Ref.~\cite{HC,HC1}. This is reflected in the operator~(\ref{opo}),
which is the radial Laplacean on symmetric spaces with zero
curvature\cite{gin62}, more precisely on the spaces of symmetric,
Hermitean, and Hermitean selfdual matrices for $\beta=1,2,4$.  Only
for $\beta=2$, the Laplacean coincides with the Laplacean over the
algebra of the group ${\rm U}(N)$. This is the only case where the
spherical function is identical to a Harish--Chandra group integral
due to the vector space isomorphism of Hermitean and anti Hermitean
matrices. For arbitrary $\beta$ the eigenvalue equation~(\ref{rado})
is closely connected to models of one dimensional interacting
particles. Using the ansatz
\begin{equation}
  \Phi_N^{(\beta)}(x,k) = \frac{\Psi_N^{(\beta)}(x,k)}
  {\Delta_N^{\beta/2}(x)\Delta_N^{\beta/2}(k)}
  \label{ansob}
\end{equation}
the eigenvalue equation~(\ref{rado}) is reduced to a Schr\"odinger
equation
\begin{eqnarray}
  &&\left(\sum_{n=1}^N\frac{\partial^2}{\partial x_n^2} - \beta\left(\frac{\beta}{2}-1\right)
    \sum_{n<m}\frac{1}{(x_n-x_m)^2}\right) \Psi_N^{(\beta)}(x,k) \ = 
-\left(\sum_{n=1}^Nk_n^2\right)\Psi_N^{(\beta)}(x,k) \
  , \label{calo}
\end{eqnarray}
which contains a kinetic part and a distance dependent interaction.
Often, one adds $N$ confining potentials to the interaction in
Eq.~(\ref{calo}).  This is done to make the system a bound state
problem. However, apart from this, the structure of the model is not
significantly affected by this modification. Thus, we will not work
with confining potentials in the sequel. The specific model
Eq.~(\ref{calo}) is also called rational CMS model \cite{serves1} or
free CMS model.

The solution $\Psi_N^{(\beta)}(x,k)$ is now interpreted as a wave
function of the Schr\"odinger equation~(\ref{calo}) with energy $\sum
k_n^2$. Thus, no symmetry condition such as
$\Psi_N^{(\beta)}(x,k)=\Psi_N^{(\beta)}(k,x)$ is imposed. In the
following we always refer to functions such as $\Psi_N^{(\beta)}(x,k)$
as wave function. On the other hand, functions such as
$\Phi_N^{(\beta)}(k,x)$ and more general solutions of eigenvalue
equations of type (\ref{rado}) are referred to as matrix Bessel
functions.

The parameter $\beta > 0$ measures the strength of the inverse
quadratic interaction. The interaction can be attractive $\beta < 2$
or repulsive $\beta > 2$.  For $\beta=2$, the model is interaction
free. This is group theoretically the unitary case and equivalent to
the Itzykson--Zuber derivation~\cite{IZ} of the ${\rm U}(N)$
Harish--Chandra integral.

The symmetric spaces mentioned above stem from a common larger group,
namely the special linear group. In Cartan´s classification they are
referred to as ${\rm A}$, ${\rm AI}$ and ${\rm AII}$ \cite{HEL}. There
are other symmetric spaces derived from the orthogonal and the
symplectic groups as larger groups, designated ${\rm B}$, ${\rm C}$
and ${\rm D}$, respectively. These symmetric spaces are also related
to Schr\"{o}dinger equations, but with a different
interaction~\cite{OP}.

\subsection{Connection to the Recursion Formula for Radial Functions}
\label{sec22}

For arbitrary positive $\beta$ the solutions of the eigenvalue
equation~(\ref{rado}) $\Phi_N^{(\beta)}(x,k)$ can be expressed in
terms of a recursion formula~\cite{GUKOP1,GUKO1}
\begin{equation}
\Phi_N^{(\beta)}(x,k) \ = \ \int d\mu(x^\prime,x) \,
         \exp\left(i\left(\sum_{n=1}^Nx_n -
         \sum_{n=1}^{N-1}x_n^\prime\right)k_N\right) \,
         \Phi_{N-1}^{(\beta)}(x^\prime,\widetilde{k}) \ ,
\label{rec1}
\end{equation}
where $\Phi_{N-1}^{(\beta)}(x^\prime,\widetilde{k})$ is the solution
of the Laplace equation~(\ref{rado}) for $N-1$.  Here, $\widetilde{k}$
denotes the set of quantum numbers $k_n, \ n=1,\ldots,(N-1)$ and
$x^\prime$ the set of integration variables $x_n^\prime, \
n=1,\ldots,(N-1)$. The integration measure is
\begin{equation}
d\mu(x^\prime,x) \ = \ G_N^{(\beta)} \, \frac{\Delta_{N-1}(x^\prime)}
                            {\Delta_N^{\beta-1}(x)} \,
                            \left(-\prod_{n,m}(x_n-x_m^\prime)\right)
                            ^{\beta/2-1} \, d[x^\prime] \ .
\label{rec2}
\end{equation}
Here, $d[x^\prime]$ is the product of all differentials $dx_n^\prime,
\ n=1,\ldots,(N-1)$. The constant $G_N^{(\beta)}$ guarantees a proper
normalization. The inequalities
\begin{equation} x_n \ \le \ x_n^\prime \ \le \ x_{n+1} \ , \qquad
  n=1,\ldots,(N-1) \label{rec3}
\end{equation}
define the domain of integration. An equivalent recursion formula
exists also for the eigenfunctions $\Psi_N^{(\beta)}(x,k)$ of the
Hamiltonian in Eq.~(\ref{calo}). For $\beta=1,2,4$ the above recursion
formula is equivalent to group integrals over ${\rm O}(N)$, ${\rm
  U}(N)$ and ${\rm USp}(2N)$, respectively. The case of arbitrary
$\beta$ has not found a clear group theoretical or geometrical
interpretation yet. However, many properties which are obvious for the
group integral carry over to arbitrary $\beta$. We just mention the
following. $\Phi_N^{(\beta)}(x,k)$ is a symmetric function in both
sets of arguments. This has as a direct consequence that the behavior
under particle exchange of the wave function $\Psi_N^{(\beta)}(x,k)$
is only governed by the Vandermonde determinant
$\Delta_N^{\beta/2}(x)\Delta_N^{\beta/2}(k)$. The wave function
obtains under particle exchange a complex phase
\begin{equation}
  P_{nm}\Psi_N^{(\beta)}(x,k)
  = \exp(-i\pi\beta/2)\Psi_N^{(\beta)}(x,k) \ ,
  \label{exchange1}
\end{equation}
with
\begin{equation}
  P_{nm}\Psi_N^{(\beta)}(x_1,\ldots,x_n,\ldots,x_m,\ldots,k)= \Psi_N^{(\beta)}(x_1,\ldots,x_m,\ldots,x_n,\ldots,k)
  \ .
  \label{exchange2}
\end{equation}
For this reason the model of Eq.~(\ref{calo}) is frequently used as
paradigm for systems with anionic statistics \cite{Ha,Wi}. A recursion
formula akin to Eq.~(\ref{rec1}) has also been derived for Jack
polynomials~\cite{OO}.

\section{Models for Interacting Particles in Superspace}
\label{sec3}

A classification of supergroups and superalgebras similar to Cartan's
classification in ordinary space can be found in
Refs.~\cite{KAC1,KAC2}. Apart from some exotic groups, there are
essentially only two families of supergroups. The general linear
supergroup ${\rm GL}(k_1/k_2)$ respectively its compact version the
unitary supergroup ${\rm U}(k_1/k_2)$ and the orthosymplectic group
${\rm OSp}(k_1/2k_2)$. A classification of the symmetric superspaces
has been given in Ref.~\cite{MRZ1}.

In Sections~\ref{sec32} and~\ref{sec33} we present supersymmetric
generalizations of models for interacting particles based on the
super\-groups ${\rm GL}(k_1/k_2)$ and on the symmetric superspaces
${\rm GL}(k_1/2k_2)/{\rm OSp}(k_1/2k_2)$. In Section~\ref{sec34}, we
give the supersymmetric generalization based on the supergroup ${\rm
  OSp}(k_1/2k_2)$. In Sections~\ref{sec331} and ~\ref{sec35} we
introduce two more general models which comprise the other models
derived before as special cases. These models can be considered as
supersymmetric generalization of the Schr\"{o}dinger
equation~(\ref{calo}) for the CMS models in ordinary space.

\subsection{Models Derived from the Superspace ${\rm GL}(k_1/k_2)$}
\label{sec32}

To extend the models in ordinary space to superspace, we begin with
models derived from the superunitary case.  The underlying symmetric
superspace is called ${\rm A|A}$ in Ref.~\cite{MRZ1}. We construct the
eigenvalue equation
\begin{equation}
\Delta_s^{({\rm u},\beta)} \lambda_{k_1k_2}^{(\beta)}(s,r) =
-\frac{1}{\sqrt{\beta}}\left(\sum_{p=1}^{k_1}r_{p1}^2+
\sum_{p=1}^{k_2}r_{p2}^2\right) \lambda_{k_1k_2}^{(\beta)}(s,r) \ .
\label{evrub}
\end{equation}
for the operator
\begin{eqnarray}
\Delta_s^{({\rm u},\beta)} &=& \frac{1}{\sqrt{\beta}}\sum_{p=1}^{k_1}
     \frac{1}{B_{k_1k_2}^\beta(s)} \frac{\partial}{\partial
     s_{p1}}B_{k_1k_2}^\beta(s) \frac{\partial}{\partial
     s_{p1}}+
     \frac{1}{\sqrt{\beta}}\sum_{p=1}^{k_2}
     \frac{1}{B_{k_1k_2}^\beta(s)} \frac{\partial}{\partial
     s_{p2}}B_{k_1k_2}^\beta(s) \frac{\partial}{\partial s_{p2}} \ ,
\label{lapub}
\end{eqnarray}
where the function~\cite{TG,GGT}
\begin{equation}
B_{k_1k_2}(s) = \frac{\prod_{p<q}(s_{p1}-s_{q1})
                      \prod_{p<q}(s_{p2}-s_{q2})}
                      {\prod_{p,q}(s_{p1}-is_{q2})}
\label{vds}
\end{equation}
is the square root of the Berezinian for the superalgebra ${\rm
  u}(k_1/k_2)$.  Using the ansatz
\begin{equation}
\lambda_{k_1k_2}^{(\beta)}(s,r) = \frac{\eta_{k_1k_2}^{(\beta)}(s,r)}
     {B_{k_1k_2}^{\beta/2}(s)B_{k_1k_2}^{\beta/2}(r)}
\label{ansub}
\end{equation}
leads to the Schr\"odinger equation
\begin{eqnarray}
&&\left(\sum_{p=1}^{k_1}\frac{\partial^2}{\partial s_{p1}^2}+
        \sum_{q=1}^{k_2}\frac{\partial^2}{\partial s_{q1}^2} -
        \beta\left(\frac{\beta}{2}-1\right)
        \sum_{p<q}\frac{1}{(s_{p1}-s_{q1})^2}
        -\beta\left(\frac{\beta}{2}-1\right)
        \sum_{p<q}\frac{1}{(s_{p2}-s_{q2})^2} \right)
        \eta_{k_1k_2}^{(\beta)}(s,r) \nonumber\\
        &&\qquad\qquad\qquad\qquad\qquad\qquad\qquad\qquad\qquad=
        -\left(\sum_{p=1}^{k_1}r_{p1}^2+\sum_{p=1}^{k_2}r_{p2}^2\right)
        \eta_{k_1k_2}^{(\beta)}(s,r) \ ,
\label{evob}
\end{eqnarray}
which includes the eigenvalue equation~(\ref{rado}) as special case
for $k_1=0$ or $k_2=0$. Again, the case $\beta=2$ gives, for all $k_1$
and $k_2$, an interaction free model, connecting to the supersymmetric
Harish--Chandra integral for the unitary supergroup ${\rm U}(k_1/k_2)$.

\subsection{Models Derived from the Symmetric Superspaces ${\rm
          GL}(k_1/2k_2)/{\rm OSp}(k_1/2k_2)$}
\label{sec33}

Also the two forms of the symmetric superspace ${\rm GL}(k_1/2k_2)/{\rm
OSp}(k_1/2k_2)$ yield new supersymmetric models as well. These spaces
are denoted ${\rm AI|AII}$ and ${\rm AII|AI}$ in
Ref.~\cite{MRZ1}. They involve the Berezinians
$\widetilde{B}_{k_12k_2}^{(c)}(s)$, see Ref.~\cite{GUH4}. Apart from
some absolute value signs which are not important here, one has $c=+i$
for the symmetric superspace ${\rm AI|AII}$ and
\begin{equation}
\widetilde{B}_{k_12k_2}^{(+1)}(s) = \frac{\prod_{p<q}(s_{p1}-s_{q1})
          \prod_{p<q}(s_{p2}-s_{q2})^4} {\prod_{p,q}(s_{p1}-is_{q2})^2}
\label{cp1}
\end{equation}
while one has $c=-i$ for the symmetric superspace ${\rm AII|AI}$ and
\begin{equation}
\widetilde{B}_{k_12k_2}^{(-1)}(s) = \frac{\prod_{p<q}(s_{p1}-s_{q1})
          \prod_{p<q}(s_{p2}-s_{q2})^4} {\prod_{p,q}(s_{p1}+is_{q2})^2} \ .
\label{cp2}
\end{equation}
Thus, we obtain the radial part of the Laplace--Beltrami operator
\begin{eqnarray}
\Delta_s^{(c)} &=& \sum_{p=1}^{k_1}
     \frac{1}{\widetilde{B}_{k_12k_2}^{(c)}(s)}
     \frac{\partial}{\partial s_{p1}} \widetilde{B}_{k_12k_2}^{(c)}(s)
     \frac{\partial}{\partial s_{p1}} + \frac{1}{2}\sum_{p=1}^{k_2}
     \frac{1}{\widetilde{B}_{k_12k_2}^{(c)}(s)}
     \frac{\partial}{\partial s_{p2}} \widetilde{B}_{k_12k_2}^{(c)}(s)
     \frac{\partial}{\partial s_{p2}} 
\label{cop}
\end{eqnarray}
and the eigenvalue equation corresponding to Eq.~(\ref{evrub}),
\begin{equation}
\Delta_s^{(c)} \rho_{k_1k_2}^{(c)}(s,r) =
-\left(\sum_{p=1}^{k_1}r_{p1}^2+\frac{1}{2}\sum_{p=1}^{k_2}r_{p2}^2\right)
\rho_{k_1k_2}^{(c)}(s,r) \ .
\label{cope}
\end{equation}
Employing the ansatz
\begin{equation}
\rho_{k_1k_2}^{(c)}(s,r) = \frac{\vartheta_{k_1k_2}^{(c)}(s,r)}
     {(\widetilde{B}_{k_12k_2}^{(c)}(s)
     \widetilde{B}_{k_12k_2}^{(c)}(r))^{1/2}} \ ,
\label{copa}
\end{equation}
we find the Schr\"odinger equation
\begin{eqnarray}
&&\left(\sum_{p=1}^{k_1}\frac{\partial^2}{\partial s_{p1}^2}+
     \frac{1}{2}\sum_{p=1}^{k_2}\frac{\partial^2}{\partial s_{p2}^2}+
     \frac{1}{2} \sum_{p<q}\frac{1}{\left(s_{p1}-s_{q1}\right)^2}- 2
     \sum_{p<q}\frac{1}{\left(s_{p2}-s_{q2}\right)^2}-
     \sum_{p,q}\frac{1}{\left(s_{p1}-c
     s_{q2}\right)^2}\right) \vartheta_{k_1k_2}^{(c)}(s,r) =\nonumber\\
     &&\qquad\qquad\qquad\qquad\qquad\qquad\qquad\qquad\qquad-\left(\sum_{p=1}^{k_1}r_{p1}^2+
     \sum_{p=1}^{k_2}\frac{r_{p2}^2}{2}\right)
     \vartheta_{k_1k_2}^{(c)}(s,r) \ .
\label{cops}
\end{eqnarray}
The choices $k_2=0$ and $k_1=0$ in Eq.~(\ref{cops}) yield
Eq.~(\ref{calo}) with $\beta=4$ and $\beta=1$, respectively. For
arbitrary $k_1$ and $k_2$ the function $\rho_{k_1k_2}^{(c)}(s,r)$ is
the supersymmetric generalization of spherical functions which we
treated in a previous work~\cite{GUKOP2,GUKO2}. For $k_1/2=k_2=k$
these models are of prominent interest in random matrix theory.  The
$k$--point eigenvalue correlation functions for a random matrix
ensemble can be expressed as derivatives of a generating functional.
This generating functional obeys a diffusion equation in supermatrix
space~\cite{GUH4} similar to Dyson's Brownian motion in ordinary
matrix space~\cite{DYS1,DYS2}.  The kernel of this diffusion equation
is given by the solution of Eq.~(\ref{cops}).

\subsection{Embedding of the ${\rm GL}(k_1/k_2)$ Based Models 
            into a Larger Class of Operators}
\label{sec331}

We now embed the functions $B_{k_1k_2}(s)$ and
$\widetilde{B}_{k_12k_2}^{(\pm 1)}(s)$ of Eqs.~(\ref{vds}),
(\ref{cp1}) and (\ref{cp2}) into a larger class of functions defined by
\begin{equation}
B_{k_1k_2}^{(c,\beta_1,\beta_2)}(s) =
                      \frac{\prod_{p<q}(s_{p1}-s_{q1})^{\beta_1}
                      \prod_{p<q}(s_{p2}-s_{q2})^{\beta_2}}
                      {\prod_{p,q}(s_{p1}-cs_{q2})^{\sqrt{\beta_1\beta_2}}} \ .
\label{cb1}
\end{equation}
Here, we introduce two parameters $\beta_1$ and $\beta_2$.  This is of
crucial importance for the resulting models. They become very rich due
to this twofold dependence. We assume that these parameters are
positive, $\beta_1,\beta_2\geq 0$. The parameter $c$ can take the
values $c=\pm i$. The functions
$B_{k_1k_2}^{(c,\beta_1,\beta_2)}(s)$ induce a differential operator
\begin{eqnarray}
\Delta_s^{(c,\beta_1,\beta_2)} &=&\frac{1}{\sqrt{\beta_1}}
     \sum_{p=1}^{k_1} \frac{1}{B_{k_1k_2}^{(c,\beta_1,\beta_2)}(s)}
     \frac{\partial}{\partial s_{p1}}
     B_{k_1k_2}^{(c,\beta_1,\beta_2)}(s) \frac{\partial}{\partial
     s_{p1}}+\frac{1}{\sqrt{\beta_2}}\sum_{p=1}^{k_2}
     \frac{1}{B_{k_1k_2}^{(\beta_1,\beta_2)}(s)}
     \frac{\partial}{\partial s_{p2}}
     B_{k_1k_2}^{(c,\beta_1,\beta_2)}(s) \frac{\partial}{\partial
     s_{p2}} \ .
\label{cb2}
\end{eqnarray}
In the first quadrant of the $(\beta_1,\beta_2)$ plane
$B_{k_1k_2}^{(c,\beta_1,\beta_2)}(s)$ and therefore
$\Delta_s^{(c,\beta_1,\beta_2)}$ is analytic in $\beta_1$ and
$\beta_2$, respectively. The eigenvalue equation corresponding to
Eq.~(\ref{evrub}) reads
\begin{equation}
\Delta_s^{(c,\beta_1,\beta_2)}
\rho_{k_1k_2}^{(c,\beta_1,\beta_2)}(s,r) =
-\left(\sum_{p=1}^{k_1}\frac{r_{p1}^2}{\sqrt{\beta_1}}+
\sum_{p=1}^{k_2}\frac{r_{p2}^2}{\sqrt{\beta_2}}\right)
\rho_{k_1k_2}^{(c,\beta_1,\beta_2)}(s,r) \ .
\label{cb3}
\end{equation}
With the ansatz
\begin{equation}
\rho_{k_1k_2}^{(c,\beta_1,\beta_2)}(s,r) =
     \frac{\vartheta_{k_1k_2}^{(c,\beta_1,\beta_2)}(s,r)}
     {B_{k_1k_2}^{(c,\beta_1/2,\beta_2/2)}(s)
     B_{k_1k_2}^{(c,\beta_1/2,\beta_2/2)}(r)} 
\label{cb4}
\end{equation}
we obtain the Schr\"odinger equation
\begin{eqnarray}
&&\left(\frac{1}{\sqrt{\beta_1}}\sum_{p=1}^{k_1}
  \frac{\partial^2}{\partial s_{p1}^2}+
  \frac{1}{\sqrt{\beta_2}}\sum_{p=1}^{k_2} \frac{\partial^2}{\partial
  s_{p2}^2}-
  \sqrt{\beta_1}\left(\frac{\beta_1}{2}-1\right)
  \sum_{p<q}\frac{1}{\left(s_{p1}-s_{q1}\right)^2}-
  \sqrt{\beta_2}\left(\frac{\beta_2}{2}-1\right)
  \sum_{p<q}\frac{1}{\left(s_{p2}-s_{q2}\right)^2}\right.\qquad\qquad\nonumber\\
  &&\qquad\qquad\qquad\qquad\left.+\frac{1}{2}\left(\sqrt{\beta_1}-\sqrt{\beta_2}\right)
  \left(\frac{1}{2}\sqrt{\beta_1\beta_2}+1\right)
  \sum_{p,q}\frac{1}{\left(s_{p1}- cs_{q2}\right)^2}\right)
  \vartheta_{k_1k_2}^{(c,\beta_1,\beta_2)}(s,r) =
\nonumber\\ &&\qquad\qquad\qquad\qquad\qquad\qquad
  -\left(\sum_{p=1}^{k_1}\frac{1}{\sqrt{\beta_1}}r_{p1}^2+
  \sum_{p=1}^{k_2}\frac{1}{\sqrt{\beta_2}}r_{p2}^2\right)
  \vartheta_{k_1k_2}^{(c,\beta_1,\beta_2)}(s,r) \ .
\label{cb5}
\end{eqnarray}
In the sequel, we refer to the model~(\ref{cb5}) as
{\it superunitary model}.

The superunitary model includes the models derived from the unitary
supergroup, discussed in Section \ref{sec32} for
$\beta_1=\beta_2=\beta$. The models derived from the symmetric spaces
${\rm AI|AII}$ and ${\rm AII|AI}$ discussed in Section \ref{sec34} are
included. They result for $\beta_1=1,\ \beta_2=4$ in the case $c=i$
and for $\beta_1=4,\
\beta_2=1$ in the case $c=-i$. The solutions
$\rho_{k_1k_2}^{(c,\beta_1,\beta_2)}$ and
$\vartheta_{k_1k_2}^{(c,\beta_1,\beta_2)}$ are real analytic functions in 
$\beta_1$ and $\beta_2$. Since
$\Delta_s^{(-c,\beta_1,\beta_2)}=\Delta_s^{(c,\beta_1,\beta_2)\dagger}$
the solutions also have the symmetry
\begin{equation}
\vartheta_{k_1k_2}^{(c,\beta_1,\beta_2)*}(s_1,s_2,r)=
\vartheta_{k_1k_2}^{(-c,\beta_1,\beta_2)}(s_1,s_2,r)=
\vartheta_{k_1k_2}^{(c,\beta_1,\beta_2)}(s_1,-s_2,r)\ .
\end{equation}
We observe that only in the case $\beta_1=\beta_2=\beta$ the
interaction between the two sets of variables vanishes. If we choose
$\beta_1=0$ and $\beta_2\neq 0$, we recover the noninteracting model,
i.e.~the Harish--Chandra integral, for the variables $r_{p1}, s_{p1}$,
$p=1\ldots k_1$. Analogously, the choice $\beta_2=0$ and $\beta_1\neq
0$ yields the noninteracting model, i.e.~the Harish--Chandra integral,
for the variables $r_{p2}, s_{p2}$, $p=1\ldots k_2$. In
Eq.~(\ref{cb5}) the points $(\beta_1,\beta_2)=(0,0)$ and
$(\beta_1,\beta_2)=(2,2)$ are indistinguishable. They both yield a
completely noninteracting model in either set of variables. As
mentioned before, the point $(2,2)$ has the group theoretical
interpretation as supersymmetric Harish--Chandra integral.

The CMS models in ordinary space Eq.~(\ref{calo}) are recovered by
setting either $k_1=0$ or $k_2=0$.  For the models Eq.~(\ref{calo})
the points of even $\beta=2,4,6,\ldots$ are special \cite{etin95}.
The wavefunction $\Phi_N^{(\beta)}$ can always be written in an
asymptotic expansion akin to the Hankel expansion of Bessel
functions~\cite{ABR}. In a previous publication~\cite{GUKOP1,GUKO1} we
showed that only for even $\beta$ this asymptotic expansion terminates
after a finite number of terms.  In the present context, this property
carries over to the points $(\beta_1,\beta_2)= (2n,2n)$, $n\in N_+$,
since there the Schr\"odinger equation Eq.~(\ref{cb5}) decouples into
the sum of two independent CMS models Eq.~(\ref{calo}). It is an
intriguing and unsolved question if there are other points in the
$(\beta_1,\beta_2)$ plane with this property. We conjecture that this
property holds for an arbitrary point $(2n,2m)$, $n,m\in N_+$.

Due to the non--Hermitecity of the left hand side, the interpretation
of Eq.~(\ref{cb5}) as a Schr\"odinger equation has to be done with
some care, see Section~\ref{sec4}.

\subsection{Models Derived from the Superspace ${\rm OSp}(k_1/2k_2)$}
\label{sec34}

Furthermore, we derive another class of models by considering the
group ${\rm OSp}(k_1/2k_2)$ instead of ${\rm GL}(k_1/k_2)$. The r\^ole
of the Berezinian $B_{k_12k_2}(s)$ is taken over by the
functions~\cite{HCUOSP}
\begin{equation}
C_{k_1k_2}(s) = \frac{\prod_{p<q}(s_{p1}^2-s_{q1}^2)
                      \prod_{p<q}(s_{p2}^2-s_{q2}^2)
                      \prod_{p=1}^{k_2}s_{p2}}
                      {\prod_{p,q}(s_{p1}^2+s_{q2}^2)}
\label{bke}
\end{equation}
for even $k_1$ and by
\begin{equation}
C_{k_1k_2}(s) = \frac{\prod_{p<q}(s_{p1}^2-s_{q1}^2)
                      \prod_{p<q}(s_{p2}^2-s_{q2}^2)
                      \prod_{p=1}^{[k_1/2]}s_{p1}}
                      {\prod_{p,q}(s_{p1}^2+s_{q2}^2)}
\label{bko}
\end{equation}
for odd $k_1$. Here, we employ the notation $[k_1/2]$ for the integer
part of $k_1/2$. The two formulae differ only in the last terms of the
numerators. We define the operator
\begin{eqnarray}
\Delta_s^{({\rm uosp},\beta)} = \frac{1}{2}\sum_{p=1}^{[k_1/2]}
           \frac{1}{C_{k_12k_2}^\beta(s)} \frac{\partial}{\partial
           s_{p1}}C_{k_12k_2}^\beta(s) \frac{\partial}{\partial
           s_{p1}}+\frac{1}{2}
           \sum_{p=1}^{k_2} \frac{1}{C_{k_12k_2}^\beta(s)}
           \frac{\partial}{\partial s_{p2}}C_{k_12k_2}^\beta(s)
           \frac{\partial}{\partial s_{p2}} \ ,
\label{uospop}
\end{eqnarray}
such that we recover the supersymmetric Harish--Chandra case for
$\beta=2$, see Ref.~\cite{HCUOSP}.  We seek the
eigenfunctions $\chi_{k_12k_2}^{(\beta)}(s,r)$ of this operator,
\begin{equation}
\Delta_s^{({\rm uosp},\beta)} \chi_{k_12k_2}^{(\beta)}(s,r) = -
2\left(\sum_{p=1}^{[k_1/2]}r_{p1}^2+\sum_{p=1}^{k_2}r_{p2}^2\right)
\chi_{k_12k_2}^{(\beta)}(s,r) \ .
\label{uospope}
\end{equation}
To arrive at a Schr\"odinger equation, we make the ansatz
\begin{equation}
\chi_{k_12k_2}^{(\beta)}(s,r) = \frac{\omega_{k_12k_2}^{(\beta)}(s,r)}
     {C_{k_12k_2}^{\beta/2}(s)C_{k_12k_2}^{\beta/2}(r)} \ ,
\label{uospopa}
\end{equation}
which yields
\begin{eqnarray}
&&\left(\frac{1}{2}\sum_{p=1}^{[k_1/2]}\frac{\partial^2}{\partial
  s_{p1}^2}+ \frac{1}{2}\sum_{q=1}^{k_2}\frac{\partial^2}{\partial
  s_{q2}^2}- \frac{\beta}{2}\left(\frac{\beta}{2}-1\right)
  \left[\sum_{p<q}\frac{(2s_{p1})^2}{\left(s_{p1}^2-s_{q1}^2\right)^2}+
  \sum_{p<q}\frac{(2s_{p2})^2}{\left(s_{p2}^2-s_{q2}^2\right)^2}+
  \sum_{p=1}^{[k_1/2],k_2}\frac{1}{s_{p1,2}^2}\right]\right)
  \omega_{k_12k_2}^{(\beta)}(s,r)\nonumber\\ &&
  \qquad\qquad\qquad\qquad\qquad\qquad\qquad\qquad\qquad=
  -2\left(\sum_{p=1}^{[k_1/2]}r_{p1}^2+\sum_{p=1}^{k_2}r_{p2}^2\right)
  \omega_{k_12k_2}^{(\beta)}(s,r) \ .
\label{uospops}
\end{eqnarray}
The last sum on the left hand side of Eq.~(\ref{uospops}) extends over
the variables $s_{p2},p=1\ldots k_2$ in case of the
Berezinian~(\ref{bke}) and over $s_{p1},p=1\ldots[k_1/2]$ in case of
the Berezinian~(\ref{bko}).  Once more, we arrive at an interaction
free model for $\beta=2$, corresponding to the supersymmetric
Harish--Chandra integral over the supermanifold ${\rm
UOSp}(k_1/2k_2)$, see Ref.~\cite{HCUOSP}.  Again, as before in case of
the unitary supergroup, there is no interaction between the two sets
of variables $s_{p1}$ and $s_{p2}$. This is so for all values of
$\beta$. We notice that for arbitrary $\beta$, the model introduced
here contains two models in ordinary space which are not included in
the models of Section~\ref{sec2}. For $k_2=0$, we obtain models based
on ${\rm O}(k_1)$ and for $k_1=0$, we obtain models based on ${\rm
USp}(2k_2)$. Both were discussed in detail in Ref.~\cite{OP}.

\subsection{Embedding of the ${\rm OSp}(k_1/2k_2)$ Based Models
            into a Larger Class of Operators}
\label{sec35}

In Section~\ref{sec331}, we embedded the models of
Sections~\ref{sec32} and~\ref{sec33} into a much richer structure with
two parameters $\beta_1$ and $\beta_2$. We now perform the analogous
embedding for the ${\rm OSp}(k_1/2k_2)$ based
models~(\ref{uospops}). Here, the result is
\begin{eqnarray}
&&\left(\frac{1}{\sqrt{\beta_1}}\sum_{p=1}^{[k_1/2]}
  \frac{\partial^2}{\partial s_{p1}^2}+
  \frac{1}{\sqrt{\beta_2}}\sum_{p=1}^{k_2} \frac{\partial^2}{\partial
  s_{p2}^2}-\sqrt{\beta_1}\left(\frac{\beta_1}{2}-1\right)
  \left[\sum_{p<q}\frac{2s_{p1}^2+2s_{q1}^2}{\left(s_{p1}^2-s_{q1}^2\right)^2}+
  l\sum_{n=1}^{[k_1/2]}\frac{1}{2s_{n1}^2}\right]\right.\nonumber\\
  &&-\sqrt{\beta_2}\left(\frac{\beta_2}{2}-1\right)
  \left[\sum_{p<q}\frac{2s_{p2}^2+2s_{q2}^2}{\left(s_{p2}^2-s_{q2}^2\right)^2}+
  (1-l)\sum_{n=1}^{k_2}\frac{1}{2s_{n2}^2}\right]
  +\left(\sqrt{\beta_1}-\sqrt{\beta_2}\right)
  \left(\frac{1}{2}\sqrt{\beta_1\beta_2}+1\right)
  \sum_{p,q}\frac{s_{p1}^2-s_{q2}^2}{\left(s_{p1}^2+s_{q2}^2\right)^2}\nonumber\\
  &&\left.-\frac{(-1)^l}{2}\sqrt{\beta_1\beta_2}\left(\sqrt{\beta_1}-\sqrt{\beta_2}\right)
    \sum_{p,q}\frac{1}{s_{p1}^2+s_{q2}^2}\right)
  \kappa_{k_1k_2}^{(\beta_1,\beta_2)}(s,r) =\nonumber\\
  &&\qquad\qquad\qquad\qquad\qquad\qquad
  -\left(\sum_{p=1}^{[k_1/2]}\frac{1}{\sqrt{\beta_1}}r_{p1}^2+
    \sum_{p=1}^{k_2}\frac{1}{\sqrt{\beta_2}}r_{p2}^2\right)
  \kappa_{k_1k_2}^{(\beta_1,\beta_2)}(s,r) \ .
\label{ousp11}
\end{eqnarray}
We introduced the quantity $l$ with $l=0$ for $k_1$ even and $l=1$ for
$k_1$ odd. In the sequel, we refer to the model~(\ref{ousp11}) as
{\it orthosymplectic model}.

For $\beta_1=\beta_2=\beta$, Eq.~(\ref{uospops}) is recovered from the
orthosymplectic model with $\omega_{k_12k_2}^{(\beta)}(s,r)=
\kappa_{k_1k_2}^{(\beta,\beta)}(s,2r)$. The discussion of
Eq.~(\ref{ousp11}) is along the same lines as the one at the end of
Section~\ref{sec331}. For $k_1$ even --- in analogy to
the model based on the unitary supergroup --- the points
$(\beta_1,\beta_2)=(1,4)$ and $(\beta_1,\beta_2)=(4,1)$ correspond to
certain symmetric superspaces, namely to the two different forms of
the symmetric superspace ${\rm OSp}(k_1/2k_2)/{\rm
GL}((k_1/2)/k_2)$. They contain the symmetric spaces ${\rm
SO}(k_1)/{\rm SL}(k_1/2)$ and ${\rm Sp}(2k_2)/{\rm SL}(k_2)$ as
submanifolds.  In Ref.~\cite{MRZ1} they are denoted ${\rm CI|DIII}$ and
${\rm DIII|CI}$, respectively.

\section{Some Specific Solutions}
\label{sec41} 

\begin{figure}[hbt]
  \begin{center}
    \epsfig{figure=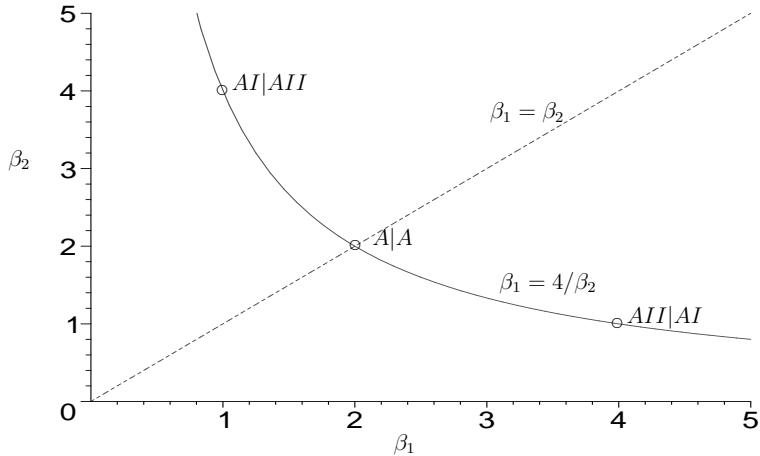,width=10cm,height=6cm,angle=0}
  \end{center}
  \caption{Curves in the $(\beta_1,\beta_2)$ plane for which we
    construct solutions. The {\em group theoretical points} are also
    indicated.}
  \label{fig3}
\end{figure}

The superunitary model~(\ref{cb5}) and the orthosymplectic
model~(\ref{ousp11}), comprising the ${\rm GL}(k_1/k_2)$ and the ${\rm
OSp}(k_1/2k_2)$ based models, respectively, have a very rich structure
due to the dependence on the two parameters $\beta_1$ and
$\beta_2$. Thus, the general solutions are highly non--trivial and not
known to us at present. Nevertheless, we are able to construct exact
solutions of the superunitary model given in Eqs.~(\ref{cb3}) and
(\ref{cb5}) for special values of the two parameters
$(\beta_1,\beta_2)$. More precisely, we derive solutions on certain
one--parameter subspaces of the $(\beta_1,\beta_2)$ plane.  We
distinguish two such one--parameter subspaces: first, the diagonal
$\beta_1=\beta_2$ and, second, the hyperbola $\beta_2=4/\beta_1$, see
Fig.~\ref{fig3}. The solutions in these subspaces contain the
solutions
of the models introduced in Sections~\ref{sec32} and~\ref{sec33}.
 
In Sections~\ref{b12} and~\ref{bh12} we state and discuss the
solutions on the diagonal and on the hyperbola, respectively. The
solutions we derive on the hyperbola are generalizations of the
recursion formula stated in Section~\ref{sec22}.  In
Section~\ref{proof} we give the derivation.  We also present some
few--particle solutions in Section~\ref{fps}.

\subsection{Solutions on the Diagonal $\beta=\beta_1=\beta_2$}
\label{b12}

In this case the Schr\"{o}dinger equation is Eq.~(\ref{evob}). It
decouples into equations for two independent sets of particles. Hence
we can write the solution as the product
\begin{equation}
\eta_{k_1k_2}^{(\beta)}(s,r)=\Psi_{k_1}^{(\beta)}(s_1,r_1)\
            \Psi_{k_2}^{(\beta)}(s_2,r_2) \ .
\label{prod1}
\end{equation}
An explicit expression of $\Psi_{k_i}^{(\beta)}(s_i,r_i),\ i=1,2$ can
be obtain in terms of the recursion formula Eq.~(\ref{rec1})
and~(\ref{rec2}) in combination with Eq.~(\ref{ansob}). We point out
that the calculation of each $\Psi_{k_i}^{(\beta)}(s_i,r_i)$
separately is in itself very difficult.

\subsection{Solutions on the Hyperbola $\beta=\beta_2=4/\beta_1$}
\label{bh12} 

In general, the interaction term between the two different sets of
one--dimensional particles at positions $s_{p1}$ and $s_{q2}$ in the
superunitary model~(\ref{cb5}) does not vanish. Thus, as the product
form of Eq.~(\ref{prod1}) is destroyed, it is highly non--trivial to
obtain solutions for positive parameters $\beta_1$ and $\beta_2$ and
for arbitrary dimensions $k_1$ and $k_2$. Nevertheless, we derive
solutions on the hyperbola $\beta_2=4/\beta_1$ shown in
Fig.~\ref{fig3}. On this hyperbola, the exponent in the denominator of
the function~(\ref{cb1}) has the constant value two. Exactly on the
same hyperbola $\beta_2=4/\beta_1$, Sergeev and Veselov constructed
supersymmetric extensions of CMS models and their solutions in terms
of deformed Jack polynomials~\cite{serves1,serves2}. The existence of
a recursion formula suggests that there should be a recursion formula
akin to the formula derived by Okounkov and Olshanski~\cite{OO} for
the deformed Jack polynomials as well.

We emphasize again that we expect the superunitary model~(\ref{cb5})
to be exactly solvable for any positive $\beta_1$,$\beta_2$.

As argued in Refs.~\cite{GUKOP1,GUKO1}, the recursion formulae
(\ref{rec1}) can be viewed as generating functions for Jack
polynomials, or equivalently, as a proper resummation. This carries
over to the present case. We generalize the supersymmetric recursion
formula of Ref.~\cite{GUKO2} for the symmetric superspace ${\rm
AI|AII}$ discussed in Section~\ref{sec34}.  We analytically continue
the solution at the point $(\beta_1,\beta_2)=(1,4)$, $(4,1)$ and
$(2,2)$. Thereby we construct the solution on the hyperbola.

We write $\rho_{k_1k_2}^{(c,\beta)}=\rho_{k_1k_2}^{(c,4/\beta,\beta)}$
for the solution of the superunitary model~(\ref{cb3}) on the
hyperbola. In Section~\ref{proof} it will be proved that
$\rho_{k_1k_2}^{(c,\beta)}$ can be expressed through the recursion
formula
\begin{eqnarray}
&&\rho_{k_1k_2}^{(c,\beta)}(s_1,s_2,r_1,r_2) \ = \int
         d\mu^{(c,\beta)}(s^\prime,s)\exp\left(i\left(\sum_{p=1}^{k_1}s_{p1} -
         \sum_{p=1}^{k_1-1}s_{p1}^\prime+\frac{\beta}{2}
         \sum_{p=1}^{k_2} |\xi_p|^2\right)r_{k_11} \right) \,
         \rho_{(k_1-1)k_2}^{(c,\beta)}(s^\prime,\widetilde{r})\quad .
\label{susyrec}
\end{eqnarray}
Here, $\rho_{(k_1-1)k_2}^{(c,\beta)}(s^\prime,\widetilde{r})$ is the
solution of the superunitary model~(\ref{cb3}) for the $k_1+k_2-1$
variables $s^\prime=(s^\prime_{11},\ldots,s^\prime_{(k_1-1)1},s^\prime_{12},\ldots,s^\prime_{k_22})$. 
The solution is labeled by the quantum numbers
$\widetilde{r}=(r_{11},\ldots,r_{(k_1-1)1},r_{12},\ldots,r_{k_22})$.
The primed variables are the integration variables. The integration
variables $s_{p1}$ are commuting. Their domain of integration is
compact and given by
\begin{equation}
  s_{p1} \ \le \ s_{p1}^\prime \ \le \ s_{(p+1)1} \ , \qquad
  p=1,\ldots,(k_1-1) \ .
  \label{ss7b}
\end{equation}
The integration variables $s_{p2}^\prime$ are related to Grassmann
variables $\xi_p$ and $\xi_p^*$ by 
\begin{equation}
|\xi_p|^2 \ = \ cs_{p2}^\prime-cs_{p2} \ .
\label{ss6}
\end{equation}
The modulus squared of a Grassmann variable is defined by
\begin{equation}
|\xi_p|^2 \ = \ \xi_p^* \xi_p \ = - \ \xi_p \xi_p^* \ ,
\label{ss6a}
\end{equation}
which is the formal analogue for the length squared of a 
commuting variable. The integration over Grassmann variables is
defined by 
\begin{equation}
\int d\xi_pd\xi_p^*=0
\quad {\rm and} \quad 
\int |\xi_p|^2 d\xi_p d\xi_p^*=1 \ .
\label{Grass}
\end{equation} 
The normalization to one differs from the convention we used in Ref.~\cite{GUKO2}, where the
integral was normalized to $1/2\pi$. The integration measure
$d\mu^{(c,\beta)}(s^\prime,s)$ reads
\begin{eqnarray}
d\mu^{(c,\beta)}(s^\prime,s)& = &\mu^{(c,\beta)}(s^\prime,s)
                                 d[\xi]d[s_1^\prime]\nonumber\\
                                 \mu^{(c,\beta)}(s^\prime,s)&=&
                                 \mu_B^{(\beta)}(s_1^\prime,s_1)\mu_F^{(c,\beta)}
                                 (s_2^\prime,s_2)\mu_{BF}^{(c,\beta)}(s^\prime,s)
                                 \ ,
\label{meas1} 
\end{eqnarray}
with the products of the differentials
\begin{equation}
d[\xi]=\prod_{p=1}^{k_2}d\xi_pd\xi_p^*
\quad {\rm and} \quad 
d[s_1^\prime]=\prod_{p=1}^{k_1-1}ds_{p1}^\prime\ , \label{ss7a}
\end{equation}
and the measure functions
\begin{eqnarray}
  \mu_B^{(\beta)}(s_1^\prime,s_1)& = &
  \Delta_{k_1}(s_1^\prime)\Delta_{k_1}^{1-4/\beta}(s_1)
  \left(-\prod_{p,q}
    \left(s_{p1}-s_{q1}^\prime\right)\right)^{2/\beta-1}
  \nonumber\\
  \mu_F^{(c,\beta)}(s_2^\prime,s_2)&=
  &\Delta_{k_2}^{\beta^2/4}(cs_2^\prime)
  \Delta_{k_2}^{\beta^2/4-\beta}(cs_2) \prod_{p\neq
    q}^{k_2} (cs_{p2}-cs_{q2}^\prime)^{\beta/2-\beta^2/4}
  \nonumber\\
  \mu_{BF}^{(c,\beta)}(s^\prime,s)& = &
  \prod_{p=1}^{k_1}\prod_{l=1}^{k_2}
  \prod_{q=1}^{k_1-1} (cs_{l2}-s_{p1})^{2-\beta/2}
  (cs_{l2}-s_{q1}^\prime)^{\beta/2-1}
  (cs_{l2}^\prime-s_{p1})^{\beta/2-1}
  (cs_{l2}^\prime-s_{q1}^\prime)^{-\beta/2}\ .
  \label{ss7}
\end{eqnarray}
We split the measure function into three parts $\mu_B,\ \mu_F,\
\mu_{BF}$ as in Ref.~\cite{GUKO2}. We do so, because the coordinates
are originally, for certain values of $\beta_1$ and $\beta_2$, 
Bosonic and Fermionic
eigenvalues of some supermatrices. The recursion
formula~(\ref{susyrec}) reproduces the recursion formula derived in
Refs.~\cite{GUKOP2,GUKO2} for $\beta=4$. It also reproduces the
supersymmetric Harish--Chandra integral discussed in Ref.~\cite{GGT}
for $\beta=2$. Moreover, for $k_2=0$ the recursion formula in ordinary
space found in Refs.~\cite{GUKOP1,GUKO1} and briefly discussed in
Section~\ref{sec22} is naturally recovered.

The case $k_1=0$ deserves some special attention, because $\mu_B$ and
$\mu_{BF}$ vanish and so does the exponential in Eq.~(\ref{susyrec}).
Importantly, the function $\mu_F$ does not.  The corresponding
Schr\"odinger equation is just that of the CMS--Hamiltonian for $k_2$
particles as defined in Eq.~(\ref{calo}). Its solution, or more
precisely the solution of its associated Laplace
equation~(\ref{rado}), is by definition given by
$\rho_{0k_2}^{(c,\beta)}=\Phi_{k_2}^{(\beta)}$.  However, the
recursion formula yields another solution
\begin{equation}
\widetilde{\Phi}_{k_2}^{(\beta)}(s_2,r_2)=\int
             d[\xi]\mu_F(s_2,s_2^\prime)
             \Phi_{k_2}^{(\beta)}(s^\prime_2,r_2) \ .
\label{altsol}
\end{equation}
For this to hold the Laplacean
$\Delta_{s_2}^{(\beta)}$ defined in Eq.~(\ref{opo}) has to commute
with the Grassmann integration of Eq.~(\ref{altsol}).  This implies
that the eigenvalues of the operator defined through the Grassmann
integration are conserved quantities. Indeed, the operator 
$\Delta_{s_2}^{(\beta)}$ commutes with the
Grassmann integral Eq.~(\ref{altsol}),
\begin{equation}
\Delta_{s_2}^{(\beta)} \int d[\xi]\mu_F(s_2,s_2^\prime)f(s^\prime_2)=
\int d[\xi]\mu_F(s_2,s_2^\prime)\Delta_{s^\prime_2}^{(\beta)}
f(s^\prime_2) \ ,
\label{commut}
\end{equation}
where $f(s^\prime_2)$ is analytic and symmetric in its arguments, but
otherwise an arbitrary test function. We sketch the derivation of
Eq.~(\ref{commut}) in~\ref{AppA}.

\subsection{Proof of the Recursion Formula}
\label{proof}

We now prove that the functions $\rho_{k_1k_2}^{(c,\beta)}$ given by
Eq.~(\ref{susyrec}) indeed solve the differential equation
Eq.~(\ref{cb2}) on the hyperbola. The proof relies on the invariance
properties of the measure function $\mu^{(c,\beta)}(s^\prime,s)$. We
define the Laplace operator Eq.~(\ref{cb2}) on the hyperbola
$\Delta_s^{(c,\beta)}=\Delta_s^{(c,4/\beta,\beta)}$ and the center of
mass momentum operator
\begin{equation}
P_s^{(c)}=\sum_{p=1}^{k_1}\frac{\partial}{\partial s_{p1}}
           - c\sum_{p=1}^{k_2}\frac{\partial}{\partial s_{p2}}
\ . \label{com}
\end{equation}
We then have the two identities
\begin{eqnarray}
  P_s^{(c)} \int d\mu(s,s^\prime) f(s_1^\prime,s_2^\prime)&=& \int
  d\mu(s,s^\prime) P_{s^\prime}^{(c)} f(s_1^\prime,s_2^\prime)
  \nonumber\\
  \Delta_s^{(c,\beta)} \int d\mu(s,s^\prime)
  f(s^\prime_1,s^\prime_2)&=& \int d\mu(s,s^\prime)
  \Delta_{s^\prime}^{(c,\beta)} f(s_1^\prime,s_2^\prime) \ ,
  \label{invariant1}
\end{eqnarray}
which hold for an arbitrary function $f(s_1,s_2)$ symmetric in both
sets of arguments $s_{p1},\ p=1\ldots k_1$ and $s_{p2},\ p=1\ldots
k_2$.  We derive Eqs.~(\ref{invariant1}) by direct calculation, using
repeated integration by part. This procedure is relatively simple for
the first equation of~(\ref{invariant1}). However, for the second one
it becomes rather tedious due to the complexity of the measure
function. Some of the steps are sketched in~\ref{AppB}. A more elegant
proof is likely to exist.

Employing the properties~(\ref{invariant1}), we can now prove the
recursion formula by acting from the left with $\Delta_s^{(c,\beta)}$
on both sides of Eq.~(\ref{susyrec}). We set
\begin{equation}
f(s_1^\prime,s_2^\prime)=\exp\left(-i\left(\sum_{p=1}^{k_1-1}s_{p1}^\prime
-\frac{\beta}{2} \sum_{p=1}^{k_2}is_{p2}^\prime\right)r_{k_11} \right)
\, \rho_{(k_1-1)k_2}^{(c,\beta)}(s^\prime,\widetilde{r}) \ ,
\label{set}
\end{equation}
and obtain
straightforwardly from~(\ref{invariant1})
\begin{eqnarray}
&&\Delta_s^{(c,\beta)}\rho_{k_1k_2}^{(c,\beta)}(s,r) \ = \int
         d\mu^{(c,\beta)}(s^\prime,s)\exp\left[i\left(\sum_{p=1}^{k_1}s_{p1} -
         \sum_{p=1}^{k_1-1}s_{p1}^\prime+\frac{\beta}{2}\sum_{p=1}^{k_2}
         |\xi_p|^2\right)r_{k_11} \right]\nonumber\\ 
	 &&\qquad\qquad\qquad\qquad
         \left(-\frac{\sqrt{\beta}}{2}r_{k_11}^2+
               \Delta_{s^\prime}^{(c,\beta)}\right)
         \rho_{(k_1-1)k_2}^{(c,\beta)}(s^\prime,\widetilde{r}) \ .
\label{proof1}
\end{eqnarray}
Since by definition we have
\begin{equation}
\Delta_{s^\prime}^{(c,\beta)}
 \rho_{(k_1-1)k_2}^{(c,\beta)}(s^\prime,\widetilde{r})=
-\left(\sum_{p=1}^{k_1-1}\frac{\sqrt{\beta}}{2}r_{p1}^2+
\sum_{p=1}^{k_2}\frac{r_{p2}^2}{\sqrt{\beta}}\right)
\rho_{(k_1-1)k_2}^{(c,\beta)}(s^\prime,\widetilde{r}) \ ,
\label{proof1a}
\end{equation}
we arrive at
\begin{equation}
\Delta_s^{(c,\beta)}\rho_{k_1k_2}^{(c,\beta)}(s,r)=
-\left(\sum_{p=1}^{k_1}\frac{\sqrt{\beta}}{2}r_{p1}^2+\sum_{p=1}^{k_2}
\frac{r_{p2}^2}{\sqrt{\beta}}\right)\rho_{k_1k_2}^{(c,\beta)}(s,r) \ ,
\label{proof2}
\end{equation}
which is our assertion.

\subsection{Few Particle Solutions}
\label{fps} 

Once the eigenfunction $\Phi_{k_2}^{(c,\beta)}(s_2,r_2)$ in ordinary
space is known, we can recursively construct the eigenfunctions
$\rho_{k_1k_2}^{(c,\beta)}(s,r)$ from formula~(\ref{susyrec}) by
starting with
$\rho_{0k_2}^{(c,\beta)}(s,r)=\Phi_{k_2}^{(c,\beta)}(s_2,r_2)$.  The
eigenfunctions $\Phi_{k_2}^{(c,\beta)}(s_2,r_2)$ are given by the
recursion formula in ordinary space, see Section~\ref{sec22}. We
illustrate the procedure for two examples in superspace.  For the sake of
simplicity, we consider only $c=+i$ and suppress the upper index $(c)$
in the sequel.

To begin with, we study the case $k_1=k_2=1$. The eigenvalue equation
is
\begin{eqnarray}
&&\left[\frac{1}{\sqrt{\beta_1}}\frac{\partial^2}{\partial s_{11}^2}+
\frac{1}{\sqrt{\beta_2}}\frac{\partial^2}{\partial
s_{12}^2}-
\frac{1}{s_{11}-is_{12}}\left(\sqrt{\beta_1}\frac{\partial}{\partial
s_{11}}+ i\sqrt{\beta_2}\frac{\partial}{\partial s_{21}}\right)\right]
\rho_{11}^{(\beta_1,\beta_2)}(s,r) \ = \ \nonumber\\
&&\qquad\qquad\qquad\qquad\qquad\qquad\qquad\qquad\qquad\qquad\qquad\qquad
-\left(\frac{r_{11}^2}{\sqrt{\beta_1}}+\frac{r_{12}^2}{\sqrt{\beta_2}}\right)
\rho_{11}^{(\beta_1,\beta_2)}(s,r)
\label{einseins}
\end{eqnarray}
yielding the closed solution
\begin{eqnarray}
\rho_{11}^{(\beta_1,\beta_2)}(s,r)&=&
           \exp\left[\pm\frac{i}{\sqrt{\beta_1}-\sqrt{\beta_2}}
         \left(\sqrt{\beta_1}s_{11}-i\sqrt{\beta_2}s_{12}\right)
          \left(r_{11}-ir_{12}\right)
         \right]
                 |\sqrt{\beta_1}-\sqrt{\beta_2}|^{\sqrt{\beta_1\beta_2}/2} z^\nu {\bf
         H}^{\mp}_{\nu}(z) \ .
\label{solu11}
\end{eqnarray}
Here, ${\bf H}_\nu(z)$ is the Hankel function of order
$\nu=\sqrt{\beta_1\beta_2/4}+1/2$. Its argument is the dimensionless
complex variable
\begin{equation}
z=\frac{\sqrt{\beta_2}
         r_{11}-i\sqrt{\beta_1}r_{12}}{\sqrt{\beta_2}-\sqrt{\beta_1}}
         \left(s_{11}-is_{12}\right) \ .
\label{wdefinition}
\end{equation}
The result~(\ref{solu11}) holds for all arbitrary positive parameters
$\beta_1$ and $\beta_2$.

From Eq.~(\ref{solu11}) we can gain deeper insight into the structure
of the solutions on the hyperbola $\beta_1\beta_2=4$. The order $\nu$
of the Hankel function becomes $3/2$ on the hyperbola. The asymptotic
Hankel series of the half integer Hankel function of order $n+1/2$
terminates after the $n$--th step~\cite{ABR}. On the other hand, the
asymptotic series of a Hankel function whose order is not
half--integer is infinite.  Thus, only the Hankel functions of
half--integer order can be expressed as a product of a finite
polynomial and an exponential.  The value $\nu=1/2$ corresponds to
either $\beta_1=0$ or $\beta_2=0$ and hence to a
one--type--of--particle model, see Eq.~(\ref{rado}) and
Eq.~(\ref{calo}). Consequently, the order $\nu=3/2$ is the lowest half
integer order describing a two--type--particle model that has a
non--trivial solution which can be written as product of a polynomial
and an exponential. Furthermore, we notice that it is exactly this
extra term in the Hankel expansion of ${\bf H}^{\mp}_{3/2}(z)$ which
can be expressed by an integration over properly chosen Grassmann
variables. Indeed the recursion formula yields directly
\begin{eqnarray}
  &&\rho_{11}^{(\beta)}\left(s_{11},s_{12},r_{11},\beta r_{12}/2\right)=
  \exp\left(ir_{11}s_{11}+i\beta r_{12}s_{12}/2\right)
\left[ \left(\frac{\beta}{2}-1\right)+
    \frac{i\beta}{2}\left(is_{12}-s_{11}\right)
\left(ir_{12}-r_{11}\right)\right]\ , \label{rec11}
\end{eqnarray}
which is identical to Eq.~(\ref{solu11}) on the hyperbola. We
expect recursive solutions of the $k_1+k_2$ particle Hamiltonian
Eq.~(\ref{cb5}) akin to the recursion formula Eq.~(\ref{susyrec}) to
exist for other half--integer $\nu$ as well.

The next simplest case is $k_1=1$ and $k_2=2$ and vice versa. It is
still possible although cumbersome to find an exact solution for
arbitrary $\beta_1$ and $\beta_2$.  As we only wish to illustrate how
the recursion works, we do not derive this exact solution here. Rather
we use formula (\ref{susyrec}) to find a solution on the hyperbola.
Without loss of generality we choose $k_2=2$ and $k_1=1$. The bosonic
measure $\mu_B(s_1,s_1^\prime)$ vanishes. We have only to perform four
Grassmann integrations. This implies that the solution can be written
as a differential operator acting on $\rho_{02}^{(c,\beta)}(s_2,\beta
r_2/2)=\Phi_{2}^{(\beta)}(s_2,\beta r_2/2)$
\begin{equation}
  \rho_{12}^{(\beta)}(s_1,s_2,r_1,\beta r_2/2)=L^{(\beta)}(s,r)\Phi_{2}^{(\beta)}(s_2,\beta r_2/2) \ .
  \label{rec21a}
\end{equation}
Using the definitions of the measure Eq.~(\ref{meas1}) and
Eq.~(\ref{ss7}) and doing the Grassmann integrations we find
\begin{eqnarray}
L^{(\beta)}(s,r)&=&\prod_{p=1}^2\left(is_{p2}-s_{11}\right)
  \left\{\prod_{q=1}^2\left[\left(\frac{\beta}{2}-1\right)
      \frac{1}{is_{q2}-s_{11}}+i\frac{\beta}{2}r_{11}-
      i\frac{\partial}{\partial
        s_{q2}}\right]\right.\nonumber\\ &
  &\quad\left.+\frac{\beta}{2}\left(\frac{\beta}{2}-1\right)
    \frac{1}{\prod_{p=1}^2\left(is_{p2}-s_{11}\right)}+
    \frac{\beta}{2}\frac{1}{\left(s_{12}-s_{22}\right)}
    \left(\frac{\partial}{\partial s_{12}}-
      \frac{\partial}{\partial s_{11}}\right)\right\}\ .
  \label{rec21b}
\end{eqnarray}
For the eigenfunction $\Phi_{2}^{(\beta)}(s_2,\beta r_2/2)$, we employ
the explicit form~\cite{GUKOP1,GUKO1}
\begin{equation}
\Phi_{2}^{(\beta)}(s_2,\beta r_2/2)\ = \
   \exp\left(-i\beta\frac{(s_{12}+s_{22})(r_{12}+r_{22})}{4}\right) \,
   \chi^{(\beta+1)}\left(\frac{\beta z}{4}\right)
\label{rec21c}
\end{equation} 
which involves the spherical functions
\begin{equation}
\chi^{(\beta+1)}(w) =2^{(\beta-1)/2}\Gamma((\beta+1)/2) 
                \frac{J_{(\beta-1)/2}(w)}{w^{(\beta-1)/2}}\ ,
\label{rec21z}
\end{equation}
where $J_\nu$ is the Bessel function of order $\nu$. The
variable $z=(s_{12}-s_{22})(r_{12}-r_{22})$ in Eq.~(\ref{rec21c}) is
dimensionless. Plugging this expression into Eq.~(\ref{rec21a}) and
using Eq.~(\ref{rec21b}) we can cast $\rho_{12}^{(\beta)}$ into the
form
\begin{eqnarray}
 \rho_{12}^{(\beta)}\left(s_{11},s_2,r_{11},\beta r_2/2\right)&=&
  \exp\left(ir_{11}s_{11}+\frac{\beta}{4}i\left(r_{12}+r_{22}\right)
    \left(s_{12}+s_{22}\right)\right)\nonumber\\
  &&\left\{\left(\frac{\beta}{2}-1\right) \left[\beta-1+
     2 z\frac{d}{dz} + i\frac{\beta}{2}\left(r_{11}-\frac{ir_{12}}{2}-\frac{ir_{22}}{2}\right)
    \left(s_{11}-\frac{is_{12}}{2}-\frac{is_{22}}{2}\right)\right]
\right.\nonumber\\
  &&\qquad\left.-\frac{\beta^2}{4}\prod_{p=1}^2
         (r_{11}-ir_{p2})(s_{11}-is_{p2})\right\}
  \chi^{(\beta+1)}\left(\beta z/4\right)\ , \label{rec21}
\end{eqnarray}
which explicitly shows the symmetry between the two sets of arguments
$s$ and $r$.

\section{Physical Interpretation}
\label{sec4}

To develop an intuition for the physics of the differential
operators~(\ref{cb5}) and (\ref{ousp11}) in superspace, we recall the
physical interpretation of CMS models in ordinary space. The
Schr\"odinger equation~(\ref{calo}) models a system of $N$ interacting
particles in one dimension, moving on the $x$--axis, say. The
eigenfunctions are labeled by a set of conserved quantities or,
equivalently, quantum numbers $k_n, \ n=1\ldots N$. This is tantamount
to saying that the system is exactly solvable. In the limit of
vanishing coupling, i.e.~for $\beta=2$, the quantum numbers are the
momenta of each particle. The characteristic feature of this model is
the $(x_n-x_m)^{-2}$ interaction potential. The models based on the
ordinary groups ${\rm O}(N)$ and ${\rm Sp}(2N)$ fit into the same
picture. However, the models have in this case a symmetry under point
reflections about $x=0$. Moreover, for the symplectic group and the
orthogonal group with $N$ odd, there is an additional inverse quadratic
confining or deconfining central potential~\cite{OP}.

We now show that the physical interpretation along those lines carries
over to our superspace models in a most natural way. We discuss the
superunitary model in Section~\ref{umodel} and the orthosymplectic
model in Section~\ref{osymodel}.

\subsection{Superunitary Model}
\label{umodel}

The superunitary model is given by Eq.~(\ref{cb5}). We notice that its
differential operator is not Hermitean. This leads to some ambiguity in
the interpretation of the model. The imaginary unit in the parameter
$c$ is due to a Wick--type--of rotation of the variables
$s_{p2}$. This was needed in Ref.~\cite{EFE83} to ensure convergence
of integrals over certain supermatrices. However, in our application,
there is no such convergence problem, as long as we do not go into a
thermodynamical discussion of the model. Thus, we undo the Wick
rotation by the substitution $is_{p2}\rightarrow s_{p2}, \ p=1\ldots
k_2$. We introduce the coupling constants
\begin{eqnarray}
g_{11}&=&\sqrt{\beta_1}\left(\frac{\beta_1}{2}-1\right)\nonumber\\
g_{22}&=&\sqrt{\beta_2}\left(\frac{\beta_2}{2}-1\right)\nonumber\\
g_{12}&=&\frac{1}{2}\left(\sqrt{\beta_1}-\sqrt{\beta_2}\right)
\left(\frac{1}{2}\sqrt{\beta_1\beta_2}+1\right)  
\label{PI0}
\end{eqnarray}
and the masses 
\begin{eqnarray}
m_1=\sqrt{\beta_1/4} \quad {\rm and} \quad m_2=-\sqrt{\beta_2/4} \ .
\label{mpio}
\end{eqnarray}
We notice that the mass $m_1$ is positive, while the mass $m_2$ is
negative. Introducing the momenta $\pi_{p1}=-i\partial/\partial
s_{p1}$ and $\pi_{p2}=-i\partial/\partial s_{p2}$, we eventually
obtain the Hermitean Hamiltonian
\begin{eqnarray}
  H&=& \sum_{p=1}^{k_1} \frac{\pi_{p1}^2}{2m_1} +
  \sum_{p=1}^{k_2} \frac{\pi_{p2}^2}{2m_2} +
  \sum_{p<q}\frac{g_{11}}{\left(s_{p1}-s_{q1}\right)^2}-
  \sum_{p<q}\frac{g_{22}}{\left(s_{p2}-s_{q2}\right)^2}
  -\sum_{p,q}\frac{g_{12}}{(s_{p1}-s_{q2})^2}\ ,
  \label{PI1}
\end{eqnarray}
with now canonical conjugate variables,
$[s_{ql},\pi_{pj}]=i\delta_{pq}\delta_{jl}$. In second quantized form it reads
\begin{eqnarray}
H&=&\sum_i \int dx \frac{1}{2m_i}\psi_i^\dagger(x)\nabla^2\psi_i(x)
          +\sum_{i,j}\int dx dx^\prime
          \frac{g_{ij}}{(x-x^\prime)^2}
    \psi_i^\dagger(x)\psi_j^\dagger(x^\prime)\psi_j(x^\prime)\psi_i(x) \ .
\label{second1}
\end{eqnarray}
The
Hamiltonian~(\ref{PI1}) describes a one--dimensional interacting
many--body system for two kinds of $k_1$ particles at positions
$s_{p1}, \ p=1,\ldots,k_1$ and $k_2$ particles at positions $s_{p2}, \
p=1,\ldots,k_2$ on the $s$ axis.

\begin{figure} [hbt]
  \begin{center}
    \epsfig{figure=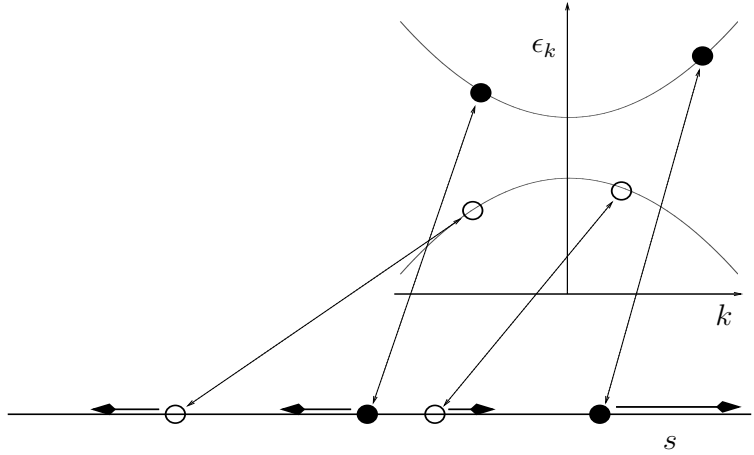,width=10cm,height=6cm, angle=0}
  \end{center}
  \caption{ Electrons in the upper (black circles) and lower (open
    circles) band of a quasi--one--dimensional semiconductor. The
    dispersion relations $\epsilon_k$ as function of the wave number
    $k$ are indicated by the parabola and the inverted parabola. The
    particles are then mapped onto the $s$ axis in the bottom part.}
  \label{fig31}
\end{figure}

The superunitary model in the form~(\ref{PI1}) may be employed to
describe electrons in a quasi--one--dimensional semiconductor, see
Fig.~\ref{fig31}. The electrons are subject to a periodic
potential. There is an upper and a lower band, separated by a gap. The
electrons in the upper band have a positive (effective) mass, while
the electrons in the lower band close to the gap have a negative
(effective) mass.  This is due to the dispersion relation $\epsilon_k$
as function of the wave number $k$. Its second derivative, i.e.~the
inverse mass, is positive in the upper, but negative in the lower
band~\cite{KIT}.  We recall that the coupling constants $g_{ij}$ are
not arbitrary, they are functions of both $\beta_1$ and
$\beta_2$. This makes it possible to model repulsive as well as
attractive interactions between equal particles and also between
different particles by choosing proper parameters $\beta_1$ and
$\beta_2$. We mention that the spectrum has to 
be bound from below by an additional
mechanism if one wants to derive thermodynamical
quantities.

\subsection{Orthosymplectic Model}
\label{osymodel}

As the orthosymplectic model (\ref{ousp11}) is derived from the
symmetric superspace ${\rm OSp}$ $(k_1/2k_2)$, it has additional
symmetries, comprising the ones found in the models based on the
ordinary groups ${\rm O}$$(N)$ and ${\rm Sp}$$(2N)$. There is a symmetry
of point reflections about $s_1=0$ and about $s_2=0$.  This renders
the differential operator of the orthosymplectic model~(\ref{ousp11})
real and thus Hermitean as it stands. It describes a
quasi--two--dimensional physical system. One set of particles at
positions $s_{p1}$ is confined to the $s_1$ axis and a second set of
particles at positions $s_{p2}$ confined to the orthogonal $s_2$
axis. As in the superunitary model, all particles interact through a
distance dependent, inverse quadratic potential. The point reflection
symmetry about the two axes implies that each particle at the position
$s_{pj}$ $>0$ with the momentum $\pi_{pj}$ has a counterpart at the
position $-s_{i1}$ with the momentum $-\pi_{pj}$. Moreover, due to the
reflection symmetry, the particles are also subjected to a confining
or deconfining inverse quadratic central potential. This generalizes
the situation described by the models from the ordinary groups ${\rm
O}(N)$ and ${\rm Sp}(N)$~\cite{OP}.

However, the orthosymplectic model has yet another important feature.
Closer inspection reveals that the potentials also contain angular
dependent terms. We now show that these are dipole--dipole
interactions, referred to as tensor forces in nuclear
physics~\cite{BOMO1}.  The general form of such a dipole--dipole
interaction in $d$ dimensions reads
\begin{equation}
V({\vec r}_p,{\vec r}_q)=
    v(|{\vec r}_p-{\vec r}_q|)\left( ({\vec e}_{pq}\cdot{\vec \sigma_p})
             ({\vec e}_{pq}\cdot{\vec \sigma_q})- \frac{1}{d}{\vec
             \sigma_p}\cdot{\vec \sigma_q}\right) \ ,
\label{tensor}
\end{equation}
where ${\vec r}_p$ is the position of particle $p$ and ${\vec
  \sigma_p}$ the dipole vector attached to it. The vector ${\vec
  e}_{pq}$ is the unit vector pointing in the direction ${\vec
  r}_p-{\vec r}_q$. The potential $v(r)$ depends on the distance
between the particles only. In nuclear physics, it is
short--ranged~\cite{BOMO1}, in our case the potential comes out
inverse quadratic, $v(r)=1/r^2$. In the following discussion we assume
$d=2$. This assumption is not a necessary one. Interpretations in
higher dimensions are also possible, but may be discussed elsewhere.
We notice that the functional form of the potential, when derived from
a Poisson equation, depends on the number of spatial dimensions. Thus,
one should not view the dipole--dipole interaction as stemming from a
Coulomb potential in the present two--dimensional interpretation.  For
$d=2$ we write Eq.~(\ref{tensor}) more explicitly as
\begin{equation}
V({\vec r}_p,{\vec r}_q)=\frac{\sigma_p \sigma_q} 
                              {|{\vec r}_p-{\vec r}_q|^2}{\vec e}_{pq}^{\ T}
\left[\matrix{\cos(\vartheta_p+\vartheta_q)&
\cos\vartheta_p\sin\vartheta_q\cr \sin\vartheta_p\cos\vartheta_q&
-\cos(\vartheta_p+\vartheta_q)}\right] {\vec e}_{pq} \ .
\label{tensorexplicit}
\end{equation}
with ${\vec \sigma}_p=\sigma_p (\cos\vartheta_p,\sin\vartheta_p)$.  In
our quasi--two--dimensional model, there are three possibilities
for the distance vectors ${\vec r}_p-{\vec r}_q$. Expressed in the coordinates 
$s_{p1}$ and $s_{p2}$, they read
\begin{equation}
{\vec r}_p-{\vec r}_q=\left[\matrix{\pm s_{p1}\pm s_{q1}\cr 0}\right]\ ,\
                                  \left[\matrix{0\cr \pm s_{p2}\pm
                                  s_{q2}}\right]\ ,\  \left[\matrix{\pm
                                  s_{p1}\cr \pm s_{q2}}\right]\ ,
\end{equation}
depending on which axis the particles $p$ and $q$ move. The angular
dependent interaction in Eq.~(\ref{ousp11}) can easily be cast into the
form~(\ref{tensorexplicit}).

\begin{figure}[hbt]
  \begin{center}
    \epsfig{figure=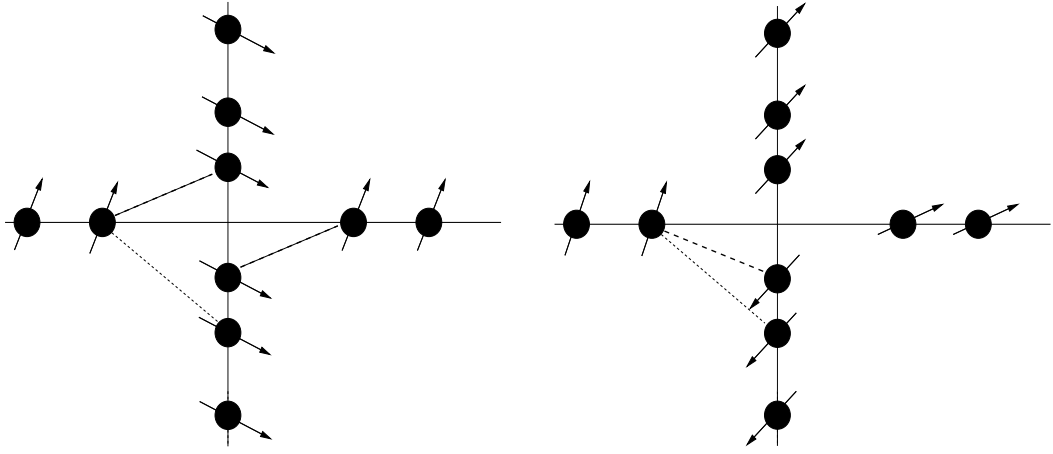,width=14cm,height=6cm,angle=0}
  \end{center}
  \caption{Two realizations of the orthosymplectic model. Left:
    $2k_1=4$ particles on the $s_1$ axis and $2k_2=6$ particles on the
    $s_2$ axis. The dipole vectors on the same axis have the same
    direction. Tensor forces are indicated as thicker and thinner
    dashed lines, corresponding to the strength of the force. The
    central forces and the distance dependent forces are not depicted.
    Right: a case with different directions of the dipole vectors on
    different sides of the same axis.} \label{fig2}
\end{figure}

Hence, the orthosymplectic model~(\ref{ousp11}) describes the motion
of two kinds of charged particles with dipole vectors attached to
them.  The interaction comprises, first, a central potential, second,
an only distance dependent potential and third a tensor force.  Two
examples are sketched in Fig.~\ref{fig2}.
Restricting ourselves to even $k_1$, we cast the Hamiltonian into the
new form
\begin{eqnarray}
2H&=& \sum_{p=1}^{k_1} \frac{\pi_{p1}^2}{2m_1}+
      \sum_{p=1}^{2k_2} \frac{\pi_{p2}^2}{2m_2} + \sum_{p\neq
    q}\frac{h_{11}}{\left(s_{p1}-s_{q1}\right)^2}
    +\sum_{p<q}\frac{h_{22}}{\left(s_{p2}-s_{q2}\right)^2}\nonumber\\ &&
    -\sum_{p,q}\frac{h_{12}}{s_{p1}^2+s_{q2}^2}+
    \sum_{p=1}^{k_1}\frac{f_1}{s_{p1}^2}
    +\sum_{p=1}^{2k_2}\frac{f_2}{s_{p2}^2}
    +\sum_{p,q}\frac{\left({\vec
    e}_{pq}\cdot{\vec \sigma_1}\right) \left({\vec
    e}_{pq}\cdot{\vec \sigma_2}\right)-{\vec
    \sigma_1}\cdot{\vec \sigma_2}/2}{s_{p1}^2+s_{q2}^2} \ ,
\label{PI2}
\end{eqnarray}
and match it on Eq.~(\ref{ousp11}) by adjusting the parameters. 
The masses are uniquely
determined. They are now both positive and given by
\begin{eqnarray}
m_1=\sqrt{\beta_1/4} \quad {\rm and} \quad m_2=\sqrt{\beta_2/4} \ .
\label{mpi2}
\end{eqnarray}
In order to determine the other free parameters in Eq.~(\ref{PI2}) 
we have to choose specific directions of the dipoles. There are various
constraints. All dipoles attached to the particles on the negative
$s_1$ axis must point into the same direction, described by the angle
$\vartheta_{1-}$, say. Similar constraints apply to the dipoles on the
other half--axes. We denote the corresponding angles by
$\vartheta_{1+}$ for the positive $s_1$ axis and with $\vartheta_{2-}$
and $\vartheta_{2+}$ for the half--axes in $s_2$
direction. Nevertheless, the four angles can not be chosen arbitrarily,
there are some further constraints which are given in~\ref{app3},
together with a complete list of all possible combinations of
different directions.  Here we only consider the possibility
$\vartheta_{1-}=\vartheta_{1+}=\vartheta_{1}$ and
$\vartheta_{2-}=\vartheta_{2+}=\vartheta_2$. Moreover, there is some
arbitrariness for choosing the moduli $\sigma_j, \ j=1,2$. For the sake
of simplicity, we assume the strengths of both dipoles to be the same
$\sigma_1=\sigma_2=\sigma$. 

The strength of the central potential in the Hamiltonian~(\ref{PI2})
is given by
\begin{eqnarray}
f_1=\frac{\beta_1}{8}\left(\frac{\beta_1}{2}-1\right) 
\quad {\rm and} \quad
f_2&=&-\frac{\beta_2}{8}\left(\frac{\beta_2}{2}-1\right)\ .
\label{coup2orth}
\end{eqnarray}
When trying to determine the coupling constants $h_{ij}$ in the
Hamiltonian~(\ref{PI2}), we face yet another type of arbitrariness.
There are at least two possibilities. The tensor force could, first,
act between pairs of particles one on either axis or it could, second,
acts between all particles. We choose the second option as it seems
more natural. The coupling constants are then given by
\begin{eqnarray}
h_{11}&=&\sqrt{\beta_1}\left(\frac{\beta_1}{2}-1\right)+
               \sigma^2\cos 2\vartheta_1\nonumber\\
h_{22}&=&\sqrt{\beta_2}\left(\frac{\beta_2}{2}-1\right)+
               \sigma^2\cos 2\vartheta_2\nonumber\\
h_{12}&=&\frac{\sqrt{\beta_1\beta_2}}{4}\left(\sqrt{\beta_1}-
         \sqrt{\beta_2}\right)\ .
\label{couporth}
\end{eqnarray} 
The strength of the dipoles is determined through the relation
\begin{equation}
\sigma^2 \cos(\vartheta_1+\vartheta_2) =
          2 \left(1+\frac{1}{2}\sqrt{\beta_1\beta_2}\right)
          \left(\sqrt{\beta_1}-\sqrt{\beta_2}\right)\ .
\label{coup3orth}
\end{equation}
A sketch of two possible realizations is given in
Fig.~\ref{fig2}.  Notice that the tensor force between two dipoles
vanishes at a relative angle of $45^\circ$ between the particle
positions.  

\begin{figure}
[hbt]
\begin{center} 
\epsfig{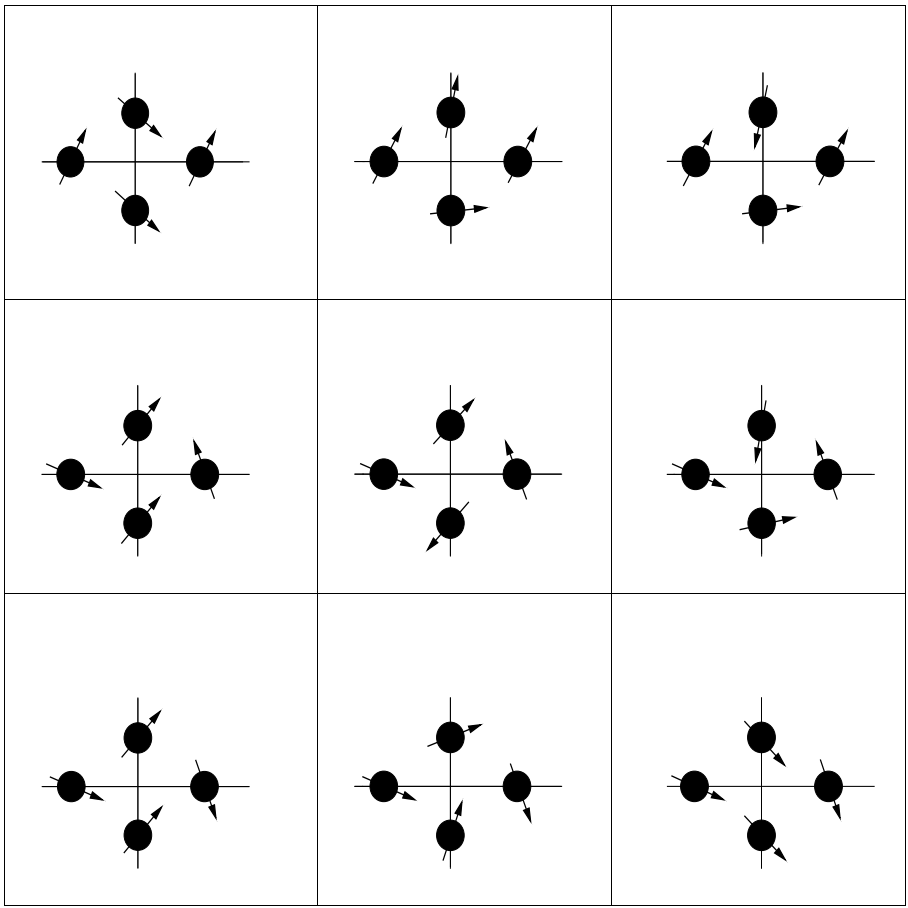}
\end{center}
\caption{Panel of the possible dipole directions in the orthosymplectic
         model as derived~\ref{app3}.}
\label{figC}
\end{figure}

Of course $H$ in Eq.~(\ref{PI2}) and the operator $\widetilde{H}$,
say, on the left hand side of Eq.~(\ref{ousp11}) are still not
identical. For $H$ and $\widetilde{H}$ to be equivalent, the time
evolution for the many--body wavefunction
$\psi_{k_1,k_2}^{(\beta_1,\beta_2)}(s,t)$ has to be the same. Thus,
the corresponding time dependent Schr\"odinger equations have to
fulfill
\begin{equation}
\label{schreq}
i\frac{\partial}{\partial t}\psi_{k_1,k_2}^{(\beta_1,\beta_2)}(s,t)\ =\
H\psi_{k_1k_2}^{(\beta_1,\beta_2)}(s,t)\ =\
\widetilde{H}\psi_{k_1k_2}^{(\beta_1,\beta_2)}(s,t) \ .
\end{equation} 
Thus, the wave function at $t=0$ must already have the reflection
symmetry
\begin{equation}
\psi_{k_1,k_2}^{(\beta_1,\beta_2)}(s,0) = \psi_{k_1,k_2}^{(\beta_1,\beta_2)}(-s,0) \qquad {\rm at} \quad t=0 \ .
\label{initial}
\end{equation}
The different interaction strengths are sketched by different widths
of the interaction lines. In~\ref{app3}, all possible combinations of
the dipole directions are derived. They are shown in Fig.~\ref{figC}.
Apart from some sign changes, all formulae given above for the
coupling constants are valid for odd $k_1$ as well.

\section{Summary and Conclusions}
\label{sec5}

Using supersymmetry, we derived new classes of models for interacting
particles. We obtained, first, a superunitary model which is based on
the supergroup ${\rm GL}(k_1/k_2)$ and on the symmetric superspace
${\rm GL}(k_1/2k_2)/{\rm OSp}(k_1/2k_2)$ and, second, an
orthosymplectic model which is based on the supergroup ${\rm
OSp}(k_1/2k_2)$. It is crucial that these models depend in a
non--trivial way on two real parameters $\beta_1$ and $\beta_2$.  Our
models extend and include the models of the CMS type in ordinary
space.

Moreover, our superunitary model contains the supersymmetric
constructions derived in Refs.~\cite{serves1,serves2}. The latter
depend on one parameter only, implying that they are defined on a
one--parameter subspace in the two--dimensional $(\beta_1,\beta_2)$
plane.  In Refs.~\cite{MEL03,MEL04a}, an ad hoc construction of models
for different kinds of particles was given, no connection to
supersymmetry was established. Not surprisingly, our superunitary
model is recovered for some parameter values in this construction. In
our approach, the connection to supersymmetry is the essential point.
It allowed us to explicity construct a complete set of solutions in
terms of recursion formulae for a trivial and a non--trivial
one--parameter subspace in the $(\beta_1,\beta_2)$ plane. This
strongly corroborates the hypothesis of exact integrability. However
an ultimate proof is still lacking. The non--trivial one--parameter
subspace coincides with the space considered in
Refs.~\cite{serves1,serves2}.  In these studies, solutions in terms of
deformed Jack polynomials were derived. The relation of the recursion
formula derived here and the deformed Jack polynomials has to be
further investigated. The recursion formulae seem to be generating
functions or, equivalently, proper resummations of the deformed Jack
polynomials. Recursion formulae on other one--parameter subspaces are
likely to exists. It would be most interesting to gain deeper insight
into the r\^ole of the one--parameter space where solutions have been
worked out. Work is in progress.

We showed that our models have a very natural interpretation.  The
superunitary model describes electrons in the upper and lower band
close to the gap in a quasi--one--dimensional semiconductor.  The
orthosymplectic model applies to a quasi--two--dimensional system of
two kinds of particles confined to two orthogonal directions. Dipole
vectors are attached to the particles. The interaction consists of
central, distance dependent and tensor forces.

\section*{Acknowledgments}

TG and HK acknowledge financial support from the Swedish Research
Council and from the RNT Network of the European Union with Grant
No.~HPRN--CT--2000-00144, respectively. HK also thanks the division of
Mathematical Physics, LTH, for its hospitality during his visits to
Lund.

\appendix

\section{Calculation of the Commutation Relation~(\ref{commut})}
\label{AppA}

Since the parameters $s_{i2}$ and the integration variables
$s_{i2}^\prime$ are related through the linear relation
of Eq.~(\ref{ss6}) we have the differentiation rules for an arbitrary
function $f$.
\begin{equation}
\frac{\partial}{\partial s_{i2}}f(s_2^\prime)=
                   \frac{\partial}{\partial
                   s_{i2}^\prime}f(s_2^\prime) ,\quad
                   \frac{\partial}{\partial s_{i2}^\prime} f(s_2)=0
                   \quad i=1\ldots k_2.
\label{A1}
\end{equation}
Acting with $\Delta_{s_2}^{(\beta)}$ onto the integral yields
\begin{eqnarray}
\Delta_{s_2}^{(\beta)} \int d[\xi]\mu_F^{(\beta)} f(s^\prime_2)=
  \int d[\xi]\left(\mu_F^{(\beta)}
  \Delta_{s^\prime_2}^{(\beta)}f(s^\prime_2)
  +f(s^\prime_2)\Delta_{s^\prime_2}^{(\beta)}\mu_F^{(\beta)}+ i\sqrt{\beta}\left(\beta-2\right)
  \mu_F^{(\beta)} \sum_{q\neq p}
  \frac{|\xi_q|^2|\xi_p|^2}{(is_{p2}-is_{q2})^3}
  \frac{\partial}{\partial s_{q2}^\prime} f(s^\prime_2)\right)\ .
\label{A2}
\end{eqnarray}
The last term in the integral has to be integrated by parts using the
rule
\begin{equation}
\int d[\xi_p]|\xi_p|^2\frac{\partial}{\partial is_{p2}^\prime}
                         f(s_2^\prime) =\int f(s_2^\prime)d[\xi_p]\ .
\label{A3}\end{equation}
We obtain
\begin{eqnarray}
&&\Delta_{s_2}^{(\beta)} \int d[\xi]\mu_F^{(\beta)} f(s^\prime_2)=\int
  d[\xi]\mu_F^{(\beta)} \Delta_{s^\prime_2}^{(\beta)}f(s^\prime_2)\
  +\nonumber\\ &&\qquad\int
  d[\xi]f(s^\prime_2)\left(-\sqrt{\beta}\left(\frac{\beta}{2}-1\right)
  \sum_{q\neq
  p}\frac{|\xi_q|^2-|\xi_p|^2}{(is_{p2}-is_{q2})^3}
  -i\sqrt{\beta}\left(\beta-2\right) \sum_{q\neq
  p} \frac{|\xi_q|^2|\xi_p|^2}{(is_{p2}-is_{q2})^3}
  \frac{\partial}{\partial s_{p2}^\prime}+
  \Delta_{s^\prime_2}^{(\beta)}\right)\mu_F^{(\beta)}\ .
\label{A4}
\end{eqnarray}
The proof is complete if the second integral vanishes identically. It
is a straightforward exercise using the definition of
$\mu_F^{(\beta)}$ in Eq.~(\ref{ss7}) and identities such as
\begin{eqnarray}
\sum_{q\neq p\neq k}\frac{|\xi_q|^2|\xi_p|^2|\xi_k|^2}
           {(is_{p2}-is_{q2})^3(is_{p2}-is_{k2})^2}&=& \sum_{q\neq
           p\neq k}\frac{|\xi_q|^2-|\xi_p|^2}
           {(is_{q2}-is_{p2})^2(is_{q2}-is_{k2})}\nonumber\\ &=&0
\label{A5}
\end{eqnarray}
to show that this is so.

\section{Derivation of the Properties~(\ref{invariant1})}
\label{AppB}

We restrict ourselves to the proof of the second equality
Eq.~(\ref{invariant1}). The proof of the first one is along the same
lines but much simpler.  It is useful to introduce the operators
\begin{eqnarray}
\Delta_{sB}^{(c,\beta)}&=&\frac{1}{\sqrt{\beta_1}} \sum_{p=1}^{k_1}
     \frac{1}{B_{k_1k_2}^{(c,\beta)}(s)} \frac{\partial}{\partial
     s_{p1}} B_{k_1k_2}^{(c,\beta)}(s) \frac{\partial}{\partial
     s_{p1}}\nonumber\\ \Delta_{sF}^{(c,\beta)}&=&
     \frac{1}{\sqrt{\beta_2}}\sum_{p=1}^{k_2}
     \frac{1}{B_{k_1k_2}^{(c,\beta)}(s)}
     \frac{\partial}{\partial s_{p2}} B_{k_1k_2}^{(c,\beta)}(s)
     \frac{\partial}{\partial s_{p2}} \ ,
\label{B1}
\end{eqnarray}
such that
\begin{equation}
\Delta_{sB}^{(c,\beta)}+\Delta_{sF}^{(c,\beta)}=\Delta_{s}^{(c,\beta)}
\ .
\label{B1a}
\end{equation}
In Eq.~(\ref{B1}) $B_{k_1k_2}^{(c,\beta)}(s)$ is the function
$B_{k_1k_2}^{(c,\beta_1,\beta_2)}(s)$ of Eq.~(\ref{cb1}) on the
hyperbola $\beta_1=4/\beta_2$. It also proves useful to split up the
the ``mixed" part of the measure function
$\mu_{BF}^{(c,\beta)}(s,s^\prime)$ as follows
\begin{eqnarray}
\mu_{BF}^{(c,\beta)}(s,s^\prime)&=&\mu_{BF1}^{(c,\beta)}(s_1,s_2,s_2^\prime)
\mu_{BF1}^{\prime(c,\beta)}(s_1^\prime,s_2,s_2^\prime) \nonumber\\
&=&\mu_{BF2}^{(c,\beta)}(s_1,s_1^\prime,s_2)
\mu_{BF2}^{\prime(c,\beta)}(s_1,s_1^\prime,s_2^\prime) \ .
\label{B2}
\end{eqnarray}
For reasons of clarity we suppress in the following the arguments of
the measure functions. Hitting the integral with
$\Delta_{sF}^{(c,\beta)}$ we get
\begin{eqnarray}
\Delta_{sF}^{(c,\beta)}\int d\mu^{(c,\beta)} f(s^\prime)&=&\int
d\mu_B^{(c,\beta)}\left(\left[\Delta_{sF}^{(c,\beta)}\mu_{BF2}^{(c,\beta)}\right]-
\mu_{BF2}^{(c,\beta)}\frac{2i}{\beta}\left(\frac{\beta}{2}-1\right)\right.\nonumber\\
&&\quad\left.\sum_{j,k}
\left(\frac{1}{is_{j2}-s_{k1}}-\frac{1}{is_{j2}-s_{k1}^\prime}\right)
\frac{\partial}{\partial s_{j2}}+\Delta_{s_2}^{(c,\beta)}\right)\int
d\mu_F^{(c,\beta)} \mu_{BF2}^{\prime(c,\beta)}f(s^\prime).
\label{B3}
\end{eqnarray}
Here and in the sequel we use the convention that operators in squared
brackets act only onto the functions inside the squared brackets. We
now pull the differential operators in $s_{p2}$, $p=1\ldots k_2$ into
the second integral using the identity Eq.~(\ref{commut}) and the
differentiation rules Eq.~(\ref{A1}). We obtain
\begin{eqnarray}
\Delta_{sF}^{(c,\beta)}\int d\mu^{(c,\beta)} f(s^\prime)&=& \int
      d\mu^{(c,\beta)}\left( \left[{\mu_{BF}^{(c,\beta)}}^{-1}
      \left(\Delta_{sF}^{(c,\beta)}+\Delta_{s_2^\prime}^{(\beta)}\right)
      \mu_{BF}^{(c,\beta)}\right] +\Delta_{s^\prime F}^{(c,\beta)}\right.
  \nonumber\\ && -
      \frac{2i}{\beta}\left(\frac{\beta}{2}-1\right)\sum_{j,k}
      \left(\frac{1}{is_{j2}-s_{k1}}-\frac{1}{is_{j2}-s_{k1}^\prime}\right)
      \left[{\mu_F^{(c,\beta)}}^{-1}\frac{\partial}{\partial s_{j2}}
      \mu_F^{(c,\beta)}+{\mu_{BF}^{(c,\beta)}}^{-1}
      \frac{\partial}{\partial
      s_{j2}^\prime}\mu_{BF}^{(c,\beta)}\right]\nonumber\\
      &&\qquad\qquad\left.
      -\frac{2i}{\beta}\left(\frac{\beta}{2}-1\right)\sum_{j,k}
      \left(\frac{|\xi_j|^2}{is_{j2}-s_{k1}}-\frac{|\xi_j|^2}{is_{j2}-s_{k1}^\prime}\right)
      \frac{\partial}{\partial s_{j2}^\prime}\right) f(s^\prime)\ .
\label{B4}
\end{eqnarray}
The last term in the right hand side has to be integrated by parts
using the rule Eq.~(\ref{A3}). Then we can write
\begin{eqnarray}
\Delta_{sF}^{(c,\beta)}\int d\mu^{(c,\beta)} f(s^\prime)&=& \int
      d\mu^{(c,\beta)}\left(\Delta_{s^\prime F}^{(c,\beta)}
      -\frac{2}{\sqrt{\beta}}
      \sum_j\left[{\mu_{BF}^{(c,\beta)}}^{-1}\frac{\partial}{\partial
      s_{j2}^\prime} \mu_{BF}^{(c,\beta)}\right]^2 \right.\nonumber\\
      &&\qquad\qquad\left.+\left[{\mu_{BF}^{(c,\beta)}}^{-1}
      \left(\Delta_{sF}^{(c,\beta)}+\Delta_{s^\prime
      F}^{(c,\beta)}\right)
      \mu_{BF}^{(c,\beta)}\right]+M_F(s,s^\prime)\right)f(s^\prime)\ ,
\label{B5}
\end{eqnarray}
where $M_F(s,s^\prime)$ is a rather lengthy expression, which contains
no further derivatives. In order to yield the calculations traceable
we state the expression explicitly
\begin{eqnarray}
M_F(s,s^\prime)&=&\sqrt{\beta}\left(\frac{\beta}{2}-1\right)
           \sum_{i\neq j\atop
           k,l}\frac{1}{is_{i2}^\prime-is_{j2}^\prime}
           \left(\frac{1}{is_{i2}^\prime-s_{k1}}-
           \frac{1}{is_{i2}^\prime-s_{l1}^\prime}\right)\nonumber\\ &&
           +\sqrt{\beta}\left(\frac{\beta}{2}-1\right)^2 \sum_{i\neq
           j\atop k,l} \left(\frac{1}{is_{i2}^\prime-s_{k1}}-
           \frac{1}{is_{i2}^\prime-s_{l1}^\prime}\right)
           \left(\frac{1}{is_{i2}^\prime-is_{j2}^\prime}-
           \frac{1}{is_{i2}^\prime-is_{j2}}\right)\nonumber\\
           &&+\sqrt{\beta}\left(\frac{\beta}{2}-1\right)^2 \sum_{i\neq
           j\atop k,l} \left(\frac{1}{is_{i2}-s_{k1}}-
           \frac{1}{is_{i2}-s_{l1}^\prime}\right)
           \left(\frac{1}{is_{i2}-is_{j2}}-
           \frac{1}{is_{i2}-is_{j2}^\prime}\right)\nonumber\\
           &&\quad-\sqrt{\beta}\left(\frac{\beta}{2}-1\right)
           \sum_{i\neq j\atop k,l}\frac{1}{is_{i2}-is_{j2}}
           \left(\frac{1}{is_{i2}-s_{k1}}-
           \frac{1}{is_{i2}-s_{l1}^\prime}\right)\nonumber\\
           &&\quad+\frac{2}{\sqrt{\beta}}\left(\frac{\beta}{2}-1\right)
           \sum_{i,k,l} \left(\frac{1}{(is_{i2}-s_{k1})^2}-
           \frac{1}{(is_{i2}-s_{l1}^\prime)^2}\right)
\label{B6}
\end{eqnarray}
For $\Delta_{sB}^{(c,\beta)}$ we proceed analogously. In this case we
use the identity Eq.~(\ref{commut}) for $\mu_B(s_1,s_1^\prime)$ and
$\Delta_{s_1}^{(4/\beta)}$ respectively, which has been proved in
Ref.~\cite{GUKO1}. We also need the integration formula (4.7) of
Ref.~\cite{GUKO2}. The outcome can be written in the same form as
Eq.~(\ref{B5})
\begin{eqnarray}
\Delta_{sB}^{(c,\beta)}\int d\mu^{(c,\beta)} f(s^\prime)&=& \int
      d\mu^{(c,\beta)}\left(\Delta_{s^\prime B}^{(c,\beta)}
      -\sqrt{\beta}
      \sum_j\left[{\mu_{BF}^{(c,\beta)}}^{-1}\frac{\partial}{\partial
      s_{j1}^\prime} \mu_{BF}^{(c,\beta)}\right]^2\right.\nonumber\\
      &&\qquad\qquad\left.+ \left[{\mu_{BF}^{(c,\beta)}}^{-1}
      \left(\Delta_{sB}^{(c,\beta)}+\Delta_{s^\prime
      B}^{(c,\beta)}\right)
      \mu_{BF}^{(c,\beta)}\right]+M_B(s,s^\prime)\right)f(s^\prime)\ .
\label{B7}
\end{eqnarray}
Here $M_B(s,s^\prime)$ is again a rather unhandy expression
\begin{eqnarray}
M_B(s,s^\prime)&=&\sqrt{\beta}\left(\frac{\beta}{2}-1\right)
           \sum_{k\neq l\atop i}\frac{1}{s_{l1}^\prime-s_{k1}^\prime}
           \left(\frac{1}{is_{i2}-s_{l1}^\prime}-
           \frac{1}{is_{i2}^\prime-s_{l1}^\prime}\right)\nonumber\\
           &&\qquad
           +\frac{2}{\sqrt{\beta}}\left(\frac{\beta}{2}-1\right)
           \left(\frac{\beta}{2}-2\right)\sum_{k\neq l\atop i}
           \frac{1}{s_{l1}-s_{k1}}
           \left(\frac{1}{is_{i2}-s_{l1}^\prime}-
           \frac{1}{is_{i2}^\prime-s_{l1}}\right)\nonumber\\ &&\
           -\frac{2}{\sqrt{\beta}}\left(\frac{\beta}{2}-1\right)^2
           \sum_{i,k,l}
           \left(\frac{1}{(is_{i2}-s_{l1}^\prime)(is_{i2}-s_{k1})}-
           \frac{1}{(is_{i2}^\prime-s_{l1}^\prime)(is_{i2}^\prime-s_{k1})}
           \right)\nonumber\\ &&\qquad
           +\sqrt{\beta}\left(\frac{\beta}{2}-1\right) \sum_{i,k}
           \left(\frac{1}{(is_{i2}-s_{k1}^\prime)^2}-
           \frac{1}{(is_{i2}^\prime-s_{k1}^\prime)^2}\right) \ .
\label{B8}
\end{eqnarray}
Adding Eqs.~(\ref{B5}) and (\ref{B7}) yields the desired result,
provided that
\begin{eqnarray}
&&M_B(s,s^\prime)+M_F(s,s^\prime)= -\left[{\mu_{BF}^{(c,\beta)}}^{-1}
\left(\Delta_{s}^{(c,\beta)}+\Delta_{s^\prime }^{(c,\beta)}\right)
\mu_{BF}^{(c,\beta)}\right]\nonumber\\ &&\ +\sqrt{\beta}
\sum_{j=1}^{k_1-1}
\left[{\mu_{BF}^{(c,\beta)}}^{-1}\frac{\partial}{\partial
s_{j1}^\prime} \mu_{BF}^{(c,\beta)}\right]^2+\frac{2}{\sqrt{\beta}}
\sum_{j=1}^{k_2}
\left[{\mu_{BF}^{(c,\beta)}}^{-1}\frac{\partial}{\partial
s_{j2}^\prime} \mu_{BF}^{(c,\beta)}\right]^2\ .
\label{B9}
\end{eqnarray}
With the definitions of $M_B$, $M_F$ in Eqs.\ (\ref{B6}), (\ref{B8})
and of $ \mu_{BF}^{(c,\beta)}$ in Eq.\ (\ref{ss7}) it is a
straightforward but extremely tedious exercise to show that
Eq. (\ref{B9}) is true.

\section{Constraints and Possible Choices for the Directions of the Dipoles
         in the Orthosymplectic Model}
\label{app3}

In order to match the right hand side of Eq.~(\ref{ousp11}) with the
Hamiltonian Eq.~(\ref{PI2}) the four angles $\vartheta_{1\pm}$,
$\vartheta_{2\pm}$ of the dipoles have to meet the following three
conditions
\begin{eqnarray}
0&=&\cos(2\vartheta_{i-})+\cos(2\vartheta_{i+})
    -2\cos(\vartheta_{i-}+\vartheta_{i+})\ ,
     \quad i=1,2 \nonumber\\
0&=&\sin(\vartheta_{1+}+\vartheta_{2-})
 +\sin(\vartheta_{1-}+\vartheta_{2+})
 -\sin(\vartheta_{1-}+\vartheta_{2-})-\sin(\vartheta_{1+}+\vartheta_{2+})\ .
\label{C1}
\end{eqnarray}
The first two equations have three solutions each. One solution is
$\vartheta_{i+}$$=\vartheta_{i-}$. The other solutions are at
$\vartheta_{i-}$ $=\frac{\pi}{2}-\vartheta_{i+}$ and at
$\vartheta_{i-}$ $=\frac{3\pi}{2}-\vartheta_{i+}$, $i=1,2$. Whenever
$\vartheta_{i+}$$=\vartheta_{i-}$ is chosen as solution the third
equation in Eq.~(\ref{C1}) does not yield any further condition. Then
two of the four angles can be chosen arbitrarily.  In general either
one or two angles can be chosen freely depending on which solution is
selected.  All possibilities are compiled in Table 1.
\begin{table}
\caption{\label{table1} The angles which can be chosen freely depending on 
the selected solution of Eq.~(\ref{C1}).}
\begin{center}
\item[]
\begin{tabular}{@{}|c||c||c||c|}\hline
 & $\vartheta_{2+}=\vartheta_{2-}$ & $\vartheta_{2+}=\frac{\pi}{2}-\vartheta_{2-}$&
$\vartheta_{2+}=\frac{3\pi}{2}-\vartheta_{2-}$ \\ \hline
$\vartheta_{1+}=\vartheta_{1-}$ & $\vartheta_{1-},\vartheta_{2-}$& 
                           $\vartheta_{1-},\vartheta_{2-}$&
                           $\vartheta_{1-},\vartheta_{2-}$\\ \hline
$\vartheta_{1+}=\frac{\pi}{2}-\vartheta_{1-}$&$\vartheta_{1-},\vartheta_{2-}$&
     $\matrix{\cr\vartheta_{1-}\ , \ (\vartheta_{2-}=\frac{\pi}{4},\frac{5\pi}{4})\cr
          \vartheta_{2-}\ , \  
  (\vartheta_{1-}=\frac{\pi}{4},\frac{5\pi}{4})\cr &}$&
          $\vartheta_{1-},\vartheta_{2-}$  \\ \hline
$\vartheta_{1+}=\frac{3\pi}{2}-\vartheta_{1-}$ &$\vartheta_{1-},\vartheta_{2-}$&
             $\vartheta_{1-},\vartheta_{2-}$&
 $\matrix{\cr\vartheta_{1-}\ , \ (\vartheta_{2-}=\frac{3\pi}{4},\frac{7\pi}{4})\cr
          \vartheta_{2-}\ , \ (\vartheta_{1-}=\frac{3\pi}{4},\frac{7\pi}{4})\cr&}$ \\ \hline
\end{tabular}\\
\end{center}
\end{table}   

The two cases depicted in Fig.~\ref{fig2} correspond to the entries
$(1,1)$ and $(2,3)$ in Table 1. In Fig.~\ref{figC} a typical
configuration for each entry of Table 1 is depicted. We also state the
general formulae for the coupling constants $h_{ij}$ and $f_i$
restricting ourselves to $k_1$ even,
\begin{eqnarray}
h_{ii}&=&\sqrt{\beta_i}\left(\frac{\beta_i}{2}-1\right)+
      \frac{\sigma_i^2}{4}
  \left(\cos(2\vartheta_{1+})+\cos(2\vartheta_{1-})+2
         \cos(2\vartheta_{1-}+\vartheta_{1+})\right)\nonumber\\
f_i&=&(-1)^i\frac{\sqrt{\beta_1}}{8}\left(\frac{\beta_1}{2}-1\right)+                                   
           \frac{\sigma_i^2}{16}\left(\cos(2\vartheta_{1+})
         +\cos(2\vartheta_{1-})+2
         \cos(2\vartheta_{1-}+\vartheta_{1+})\right)\nonumber\\
h_{12}&=&\frac{\sqrt{\beta_1\beta_2}}{4}\left(\sqrt{\beta_1}-
         \sqrt{\beta_2}\right)\ .
\label{C2}
\end{eqnarray} 
In the general case, we find
\begin{eqnarray}
\sigma_1 \sigma_2 \left(\cos(\vartheta_{1+}+\vartheta_{2+})
+\cos(\vartheta_{1+}+\vartheta_{2-})+\cos(\vartheta_{1-}+\vartheta_{2+})+
\cos(\vartheta_{1-}+\vartheta_{2-})\right) &=& 
\left(\sqrt{\beta_1-\beta_2}\right)\left(\sqrt{\beta_1\beta_2}+2\right) 
\label{C3}
\end{eqnarray} 
for the moduli squared of the dipoles.

\end{document}